\documentclass[twocolumn,amsmath,amssymb,pra,superscriptaddress,showpacs]{revtex4}
\usepackage[dvips]{graphicx}% Include figure files [dvips]
\usepackage{dcolumn}% Align table columns on decimal point
\usepackage{bm}% bold math
\newcounter{one}
\usepackage{braket}
\usepackage[tight]{subfigure}
\usepackage{color}

\newcommand{\1}{\mbox{1}\hspace{-0.25em}\mbox{l}}

\begin{document}

\title{Gain tuning for continuous-variable quantum teleportation \\ of discrete-variable states}
\author{Shuntaro Takeda}
\email{takeda@alice.t.u-tokyo.ac.jp}
\affiliation{Department of Applied Physics, School of Engineering, The University of Tokyo,\\ 7-3-1 Hongo, Bunkyo-ku, Tokyo 113-8656, Japan}
\author{Takahiro Mizuta}
\affiliation{Department of Applied Physics, School of Engineering, The University of Tokyo,\\ 7-3-1 Hongo, Bunkyo-ku, Tokyo 113-8656, Japan}
\author{Maria Fuwa}
\affiliation{Department of Applied Physics, School of Engineering, The University of Tokyo,\\ 7-3-1 Hongo, Bunkyo-ku, Tokyo 113-8656, Japan}
\author{Hidehiro Yonezawa}
\affiliation{Department of Applied Physics, School of Engineering, The University of Tokyo,\\ 7-3-1 Hongo, Bunkyo-ku, Tokyo 113-8656, Japan}
\author{Peter van Loock}
\affiliation{Institute of Physics, Johannes-Gutenberg Universit\"at Mainz, Staudingerweg 7,\\ 55128 Mainz, Germany}
\author{Akira Furusawa}
\email{akiraf@ap.t.u-tokyo.ac.jp}
\affiliation{Department of Applied Physics, School of Engineering, The University of Tokyo,\\ 7-3-1 Hongo, Bunkyo-ku, Tokyo 113-8656, Japan}

%%%%%%%%%%%%%%%%% Abstract %%%%%%%%%%%%%%%%%%%%%%%%%%%%%%%%%%%%%%%%%%%%%%%%%%%%%%%%%%%%%%%%%%%%%%%%%

\begin{abstract}
We present a general formalism to describe continuous-variable (CV) quantum teleportation of discrete-variable (DV)
states with gain tuning,
taking into account experimental imperfections.
Here the teleportation output is given by independently transforming each density matrix element of the initial state.
This formalism allows us to accurately model various teleportation experiments and to analyze the gain dependence of
their respective figures of merit.
We apply our formalism to the recent experiment of CV teleportation of qubits [S. Takeda \textit{et al.}, Nature {\bf 500}, 315 (2013)]
and investigate the optimal gain for the transfer fidelity.
We also propose and model an experiment for CV teleportation of DV entanglement.
It is shown that, provided the experimental losses are within a certain range, DV entanglement can be teleported for any non-zero squeezing by optimally tuning the gain.
\end{abstract}


%%%%%%%%%%%PACS2010%%%%%%%%%%%%%%%%%%%%%%%%%%%%%%%%%%%%%%%%%%%%%%%%%
%03.65.Wj State reconstruction, quantum tomography
%03.65.Ud Entanglement and quantum nonlocality
%03.67.Ac Quantum algorithms, protocols, and simulations 
%03.67.-a Quantum information
%03.67.Hk Quantum communication 
%03.67.Mn Entanglement measures, witnesses, and other characterizations
%42.50.Dv Quantum state engineering and measurements  
%42.50.Ex Optical implementations of quantum information processing and transfer in quantum optics
%42.65.Yj Optical parametric oscillators and amplifiers

\maketitle

%%%%%%%%%%%%%%%%% Introduction %%%%%%%%%%%%%%%%%%%%%%%%%%%%%%%%%%%%%%%%%%%%%%%%%%%%%%%%%%%%%%%%%%%%%

\section{Introduction}

Quantum teleportation~\cite{93Bennett} plays a central role in the transfer and manipulation of quantum states.
It was originally proposed for discrete-variable (DV) two-level systems~\cite{93Bennett},
and later extended to continuous-variable (CV) systems in an infinite-dimensional Hilbert space~\cite{94Vaidman,98Braunstein}.
In optics, experimental realizations have followed for both systems.
DV teleportation has been performed for qubits represented by photons,
albeit probabilistically and post-selectively~\cite{97Bouwmeester,02Lombardi}.
In contrast, CV teleportation has been performed deterministically
for quadrature variables of electromagnetic fields, however, with a
relatively low fidelity due to the finite level of resource squeezing~\cite{98Furusawa,11Lee}.
Recently, Ref.~\cite{13Takeda} reported a ``hybrid'' experiment
-- CV teleportation of qubits --
and overcame the previous limitations both in the DV and the CV regime.
Not only deterministic qubit teleportation was realized there,
but it was also demonstrated that
tuning the feedforward gain in CV teleportation
enables one to faithfully transfer qubit information
even with finite squeezing, eventually
leading to higher fidelities.

The usefulness of gain tuning for teleporting DV states is well-known.
However, a full gain optimization is a non-trivial problem.
An accurate model for the hybrid teleportation scheme is required
to obtain the optimal gain for every experimental setup and figure of merit.
Thus far, gain-tuned teleportation of qubits or single photons has been theoretically analyzed in
Refs.~\cite{99Polkinghorne,00Ralph,01Ralph,10Mista}
using the Heisenberg picture and the Wigner function.
However, these models are specifically adapted to investigate the optimal gains for certain figures of merit,
such as the value of the Clauser-Horne-type inequality,
visibility,
or the negativity of the Wigner function.
Thus, they cannot be directly applied to more general cases.
A different model, employing a more general density-matrix formalism, has been developed in Refs.~\cite{00Hof,01Hof,01Ide,02Ide}
by introducing the so-called transfer operator.
Though the transformation of DV states can be intuitively and explicitly described in this model,
these calculations assumed an ideal loss-free condition
when the input state and the resource squeezing are perfectly pure.
The extension of this model to the realistic situation including losses is hindered
by the complexity of taking into account the impurity of squeezing in the density-matrix formalism.

Here we present a general formalism to explicitly describe CV teleportation of DV states with gain tuning.
The key element of our formalism is to define the transformation of each density matrix element in the teleportation channel using Wigner functions.
The density matrix of the teleportation output is then given by independently transforming each density matrix element of the initial state.
The experimental losses in the input state and squeezing can be included in the Wigner functions,
and hence realistic experimental conditions can be simulated.

Our formalism can be straightforwardly applied to various hybrid teleportation experiments
to investigate the optimal gain tuning for a given figure of merit.
In this paper, we apply the formalism to two types of hybrid teleportation experiments.
First, the CV teleportation of photonic qubits in Ref.~\cite{13Takeda} is modeled.
We investigate the gain dependence of fidelity considering experimental inefficiencies,
and we show that our model is in good agreement with the experimental results.
Second, CV teleportation of DV entanglement is proposed and modeled.
Such an experiment can be readily implemented with current technology
and it would allow for a more efficient transfer of DV entanglement than previous teleportation schemes.
We derive a sufficient condition to teleport DV entanglement,
and we prove that, provided the experimental losses are within a certain range,
DV entanglement can be teleported for any non-zero squeezing by optimally tuning the gain.

This paper is organized as follows.
Our general formalism is derived in Secs.~\ref{sec:theory} and \ref{sec:vac_single}.
First, CV teleportation is generally modeled in the Wigner-function formalism and the effect of gain tuning is discussed in Sec.~\ref{sec:theory}.
Section~\ref{sec:vac_single} then focuses on CV teleportation of DV states, deriving the formalism to describe the
transformation of density matrices.
This formalism is applied to two specific cases in the following two sections.
Section~\ref{sec:qubit_tele} considers CV teleportation of a photonic qubit, investigating the optimal gains to achieve maximal fidelity.
In Sec.~\ref{sec:swapping}, an experiment for CV teleportation of DV entanglement is proposed and modeled based on our formalism, including a discussion on the condition for teleporting entanglement.
Finally, Sec.~\ref{sec:conclusion} concludes this paper.

%%%%%%%%%%%%%%%%% CV teleportation %%%%%%%%%%%%%%%%%%%%%%%%%%%%%%%%%%%%%%%%%%%%%%%%%%%%%%%%%%%%%%%%%%%%%

\section{Gain tuning of CV teleportation in Wigner function formalism}\label{sec:theory}

\begin{figure}[!b]
\begin{center}
\includegraphics[scale=0.6]{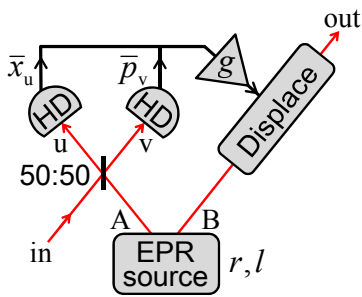}
\end{center}
\caption{(color online) Schematic of CV teleportation. HD, Homodyne detection.}
\label{fig:schematic_teleportation}
\end{figure}

Wigner functions are useful tools to describe Gaussian states and operations,
including the effect of photon losses, in a simpler way compared to density matrices.
The aim of this section is to derive an input-output relation for CV teleportation in the Wigner-function formalism,
including parameters for gain, squeezing level, and loss on the squeezing.
Our results can be regarded as a generalization of those in Refs.~\cite{98Braunstein,10Mista,03Liu},
which do not include all these parameters at the same time.
From the derived formula, we show that the effect of a non-unit-gain teleportation channel is
closely related to that of an attenuation and an amplification channel.

We start with following the standard Braunstein-Kimble protocol~\cite{98Braunstein} in the non-unit gain regime.
The schematic of CV teleportation is depicted in Fig.~\ref{fig:schematic_teleportation}.
The initial step for a sender ``Alice'' and a receiver ``Bob'' is to share
a two-mode Einstein-Podolsky-Rosen (EPR) entangled state.
It can be approximately generated by suitably mixing at a beam splitter two impure single-mode squeezed states
of squeezed quadrature variance $V_\text{sq}=[(1-l)e^{-2r}+l]/2$
and anti-squeezed quadrature variance $V_\text{an}=[(1-l)e^{2r}+l]/2$,
where $\hbar=1$, $r$ is a squeezing parameter, and $l$ denotes loss ($0\le l\le1$).
The Wigner function of the EPR state is written as a function of
quadratures of Alice's mode $\xi_\text{A} = (x_\text{A}, p_\text{A})^T$
and Bob's mode $\xi_\text{B} = (x_\text{B}, p_\text{B})^T$:
\begin{align}
W_{\text{EPR}}(\xi_\text{A},\xi_\text{B} )&= \frac{1}{4\pi^2 V_{\text{an}} V_{\text{sq}}} e^{-\xi_\text{AB} ^T \Gamma^{-1} \xi_\text{AB} }.
\end{align}
Here $\xi_\text{AB}=(x_\text{A}, p_\text{A}, x_\text{B}, p_\text{B})^T$ and
\begin{align}
\Gamma &= \begin{pmatrix}  V\1 &  C\sigma_\text{z} \\
C\sigma_\text{z} & V\1 \end{pmatrix}
\end{align}
is the covariance matrix with
$V=V_{\text{an}}+V_{\text{sq}}$,
$C=V_{\text{an}}-V_{\text{sq}}$,
$\1$ an identity matrix,
and $\sigma_\text{z}$ a Pauli matrix.
Alice then mixes her part of the EPR state and the input state
$W_{\text{in}}\left(\xi_{\text{in}}\right)$, where $\xi_{\text{in}}= (x_\text{in}, p_\text{in})^T$,
by a 50:50 beam splitter:
$(\xi_{\text{in}},\xi_{\text{A}})\to(\xi_{\text{u}},\xi_{\text{v}})
=\left((\xi_\text{in}-\xi_\text{A})/\sqrt{2},(\xi_\text{in}+\xi_\text{A})/\sqrt{2}\right)$.
The resulting overall Wigner function is
\begin{align}
W_\text{tot}(\xi_\text{u}, \xi_\text{v}, \xi_\text{B})
\!=\!W_\text{in}\!\left(\!\frac{\xi_\text{u}\!+\!\xi_\text{v}}{\sqrt{2}}\!\right)\!
W_\text{EPR}\!\left(\!\frac{\xi_\text{v}\!-\!\xi_\text{u}}{\sqrt{2}}, \xi_\text{B}\!\right),
\end{align}
where $\xi_i= (x_i, p_i)^T$ for $i=\text{u}, \text{v}$.
Alice measures $(x_\text{u}, p_\text{v})$ and sends the results $\zeta=(\overline{x}_\text{u},\overline{p}_\text{v})^T$ to Bob,
who displaces his part of the EPR state with feedforward gain $g>0$ as $\xi_\text{out} = \xi_\text{B} + \sqrt{2} g\zeta$.
The final teleported state is obtained by integrating over all possible measurement results $\zeta$~\cite{98Braunstein,03Liu},
\begin{align}
&W_\text{out} (\xi_\text{out} )\nonumber \\
&=\int d\overline{x}_\text{u}dp_\text{u}dx_\text{v}d\overline{p}_\text{v}
W_\text{tot} (\overline{x} _\text{u}, p_\text{u}, x_\text{v} ,\overline{p} _\text{v}, \xi_\text{out}-\sqrt{2} g\zeta)\nonumber\\
&= \frac{1}{g^2} \left[  W_\text{in} \circ G_{\tau} \right] \left( \frac{\xi_\text{out}}{g} \right).
\label{imperfect_EPR_GC}
\end{align}
Here $\circ$ denotes convolution, and
$G_{\tau}(\xi) = (2\pi \tau)^{-1}\exp\left[ -\xi^T\xi/(2\tau) \right]$ is a normalized Gaussian
of variance
\begin{equation}
\tau = \frac{V_\text{an}}{2} \left( 1-\frac{1}{g}   \right) ^2  + \frac{V_\text{sq}}{2} \left( 1+\frac{1}{g}   \right) ^2.
\label{gamma}
\end{equation}
Equation~(\ref{imperfect_EPR_GC}) shows that the CV teleportation channel for a given
$(r,l,g)$ is equivalent to
a thermalization process described by a Gaussian convolution
$W_\text{in}(\xi)\to [W_\text{in}\circ G_\tau](\xi)$
followed by a rescaling $\xi\to\xi/g$ in phase space.
For $l=0$, $W_\text{out}(\xi)=W_\text{in}(\xi)$ is obtained
in the limit of $r\to\infty$ at $g=1$;
otherwise the performance is limited by $l$, as is indicated by $\tau\to l$ in the limit of $r\to\infty$.
Note that for $l=0$, Eq.~(\ref{imperfect_EPR_GC}) becomes equivalent to Eq.~(6) in Ref.~\cite{03Liu},
and for $l=0$ and $g=1$, it is simplified to
$W_\text{out} (\xi_\text{out} )= [W_\text{in} \circ G_{e^{-2r}}](\xi_\text{out})$
and coincides with Eq.~(4) in Ref.~\cite{98Braunstein}.
The special case of $\xi_\text{out}=0$ in Eq.~(\ref{imperfect_EPR_GC}) was also derived and used in Ref.~\cite{10Mista}.
In this sense, the input-output relation of Eq.~(\ref{imperfect_EPR_GC}) generalizes all previous ones in the Wigner-function formalism.

The process of non-unit-gain teleportation, described in Eq.~(\ref{imperfect_EPR_GC}), can be
explained by a combination of
unit-gain teleportation, pure attenuation, and pure amplification
(``pure'' indicates the optimal attenuation or amplification with minimum excess noise~\cite{10Leonhardt}).
The pure attenuation is a channel which applies ``beam-splitter loss'' of $1-\epsilon$ ($0<\epsilon<1$) to the input state.
It can be written as
$\hat{a}_\text{out}=\sqrt{\epsilon}\hat{a}_\text{in}+\sqrt{1-\epsilon}\hat{a}_\text{vac}$ in the Heisenberg picture,
where each $\hat{a}$ denotes an annihilation operator of the output, input, and auxiliary vacuum mode, respectively.
In the Wigner-function formalism, the input-output relation is given by~\cite{10Leonhardt,11Hugo}
\begin{align}
W_\text{out}(\xi_\text{out})=\frac{1}{\epsilon}\left[W_\text{in}\circ G_\frac{1-\epsilon}{2\epsilon}\right]\left(\frac{\xi_\text{out}}{\sqrt{\epsilon}}\right).
\label{eq:att_channel}
\end{align}
In contrast, the pure-amplification channel amplifies the input signal as
$\hat{a}_\text{out}=\sqrt{\gamma}\hat{a}_\text{in}+\sqrt{\gamma-1}\hat{a}_\text{vac}^\dagger$ ($\gamma>1$),
described in the Wigner-function formalism by~\cite{10Leonhardt}
\begin{align}
W_\text{out}(\xi_\text{out})=\frac{1}{\gamma}\left[W_\text{in}\circ G_\frac{\gamma-1}{2\gamma}\right]\left(\frac{\xi_\text{out}}{\sqrt{\gamma}}\right).
\label{eq:amp_channel}
\end{align}
All Eqs.~(\ref{imperfect_EPR_GC}), (\ref{eq:att_channel}), and (\ref{eq:amp_channel}) represent Gaussian channels composed of a Gaussian convolution and phase-space rescaling.
Importantly, two successive Gaussian channels of convolution of $G_{\tau_i}$ followed by a rescaling $\xi\to\xi/g_i$ ($i=1, 2$)
can be reduced to one Gaussian channel of $G_{\tau^\prime}$ and $\xi\to\xi/g^\prime$ with $\tau^\prime=\tau_1+\tau_2/g_1^2$ and $g^\prime=g_1g_2$.
In the case of $0<g<1$,
Eq.~(\ref{imperfect_EPR_GC}) can be decomposed into two successive Gaussian channels
of $\tau_1=\tau-\tau_2$, $g_1=1$ and $\tau_2=(1-g^2)/2g^2$, $g_2=g$.
This means that below-unit-gain teleportation can be regarded as a sequence of unit-gain teleportation (convolution of $G_{\tau_1}$)
and pure attenuation [Eq.~(\ref{eq:att_channel}) at $\epsilon=g^2$].
For a given $(r,l)$,
the thermalization effect of $G_{\tau_1}$ is minimized at the gain of $g_\text{att}=(V_\text{an}+V_\text{sq}-1)/(V_\text{an}-V_\text{sq})=\tanh r$,
where $\tau_1$ takes its minimum value of $\tau_1^\text{min}=(4V_\text{an}V_\text{sq}-1)/[2(V_\text{an}+V_\text{sq}-1)]=l$.
Interestingly, this gain depends only on $r$ (not on $l$),
and the minimum $\tau_1$ is equal to the lower bound of $\tau$ in the unit-gain regime ($\tau\to l$ for $g=1$ and $r\to\infty$, as mentioned above).
When $l=0$ and thus $\tau_1=0$, the teleportation channel at $g_\text{att}$ is equivalent to a pure-attenuation channel at $\epsilon=\tanh^2r$.
Otherwise $\tau_2$ is always positive and the thermalization effect is unavoidable.
A similar discussion can be made for $g>1$.
By decomposing Eq.~(\ref{imperfect_EPR_GC}) into two successive Gaussian channels
of $\tau_1=(g^2-1)/2g^2$, $g_1=g$ and $\tau_2=g^2(\tau-\tau_1)$, $g_2=1$,
above-unit-gain teleportation can be regarded as a sequence of pure amplification [Eq.~(\ref{eq:amp_channel}) at $\gamma=g^2$]
and unit-gain teleportation (convolution of $G_{\tau_2}$).
The minimum value of $\tau_2^\text{min}=l$ is achieved at the gain of $g_\text{amp}=\tanh^{-1} r$.
For $l=0$, the teleportation channel at $g_\text{amp}$ becomes equivalent to a pure-amplification channel at $\gamma=\tanh^{-2}r$.
The equivalence to pure-attenuation and pure-amplification channels at $l=0$ has been derived using different methods in Refs.~\cite{99Polkinghorne,01Hof,99Ralph}.

Thus far, we have only discussed one simple CV teleportation channel, as expressed by Eq.~(\ref{imperfect_EPR_GC}).
However, an actual experimental situation is typically more complex.
For example, in single-mode CV teleportation experiments~\cite{98Furusawa,03Bowen,03Zhang,07Yonezawa2,12Takeda2,11Lee},
an input state is first attenuated (loss), next teleported, and finally attenuated again by the measurement (finite measurement efficiency);
the overall channel thus should be written as three consecutive Gaussian channels.
However, such a complex channel, composed of consecutive CV teleportation, pure attenuation, and amplification,
can be described by only one input-output relation, similar to Eq.~(\ref{imperfect_EPR_GC}).
The reason is that two consecutive Gaussian channels can be reduced to one Gaussian channel, as mentioned above;
thus any number of consecutive Gaussian channels can be simplified to just one, written as
\begin{align}
W_\text{out}(\xi_\text{out})=\frac{1}{g_\text{tot}^2}\left[W_\text{in}\circ G_{\tau_\text{tot}}\right]\left(\frac{\xi_\text{out}}{g_\text{tot}}\right),
\label{eq:overall_channel}
\end{align}
with only two parameters $\tau_\text{tot}$ and $g_\text{tot}$ to characterize the overall channel property.
This is of great convenience to describe various types of CV teleportation experiments in an accurate and compact fashion.
Besides the example of single-mode CV teleportation,
Eq.~(\ref{eq:overall_channel}) is also applicable to the case when CV teleportation is performed successively~\cite{07Yonezawa}.

%%%%%%%%%%%%%%%%% Qubit teleportation %%%%%%%%%%%%%%%%%%%%%%%%%%%%%%%%%%%%%%%%%%%%%%%%%%%%%%%%%%%%%%%%%%%%%

\section{Teleportation of DV state \hspace{20mm} in transition-operator formalism}\label{sec:vac_single}

The next step is to derive the transformation of DV states in a CV teleportation channel
by applying the input-output relation of Eq.~(\ref{imperfect_EPR_GC}) to such states.
Discrete-variable states of particular interest are the following two types of a qubit.
One is a singe-rail encoded qubit,
\begin{equation}
\ket{\psi}_\text{s} = \alpha  \ket{0}  + \beta \ket{1},
\label{single_rail_qubit}
\end{equation}
expressed in the photon number basis ($|\alpha|^2+|\beta|^2=1$),
where the photon number directly represents the logical ``0'' and ``1''
(see, \textit{e.g.}, Refs.~\cite{02Lombardi,02Sciarriono}).
The other is a dual-rail encoded qubit,
\begin{equation}
 \ket{\psi}_\text{d} = \alpha \ket{0}_\text{X}\otimes\ket{1}_\text{Y}+\beta \ket{1}_\text{X}\otimes\ket{0}_\text{Y},
\label{dual_rail_qubit}
\end{equation}
where the logical ``0'' and ``1'' are represented by the mode (X or Y) in which the photon is present
(see, \textit{e.g.}, Refs.~\cite{97Bouwmeester,13Takeda,98Pan}).
Continuous-variable teleportation of these states can be expressed by a transformation between non-Gaussian Wigner functions using Eq.~(\ref{imperfect_EPR_GC}).
However, below we may convert our representation from Wigner functions to density matrices
in order to describe the transformation more intuitively and conveniently.
The formalism derived below is applicable to various types of hybrid teleportation experiments.

In order to calculate the density matrix corresponding to a given Wigner function,
we start with introducing the following function of an operator $\hat{A}$:
\begin{align}
W^{\hat{A}}\!(x,p)\!=\!\frac{1}{2\pi}\!\int\!dy e^{ipy}\!\Braket{x-\frac{y}{2}|\hat{A}|x+\frac{y}{2}}.
\label{eq:Moyal_function}
\end{align}
This is a generalized version of the Wigner function for an arbitrary, not necessarily Hermitian, operator $\hat{A}$.
When $\hat{A}$ is a density operator, this function describes the Wigner function of the corresponding state.
For $\hat{A}=\ket{m}\!\bra{n}$ (photon number basis, $m \geq n \geq 0$),
Eq.~(\ref{eq:Moyal_function}) can be expressed by the Laguerre function $\mathcal{L}$~\cite{01Wolf} as
\begin{align}
W^{\ket{m}\!\bra{n}} (\xi ) =& \frac{(-1)^n}{\pi} \sqrt{\frac{n!}{m!}} \left( \sqrt{2} v^T\xi \right)^{m-n} \nonumber \\
&\quad\times \mathcal{L}^{m-n}_n \left( 2 \xi^T \xi  \right)  \exp \left[- \xi^T \xi \right],
\label{photon_num_wigner}
\end{align}
where $\xi=(x,p)^T$ and $v=(1,-i)^T$.
In the case of $n>m\ge0$, we need to replace
$n\to m$, $m\to n$ and $\xi\to\sigma_\text{z} \xi$
in the right hand side of Eq.~(\ref{photon_num_wigner}).
When an input state $\hat \rho_\text{in}$ is expanded in the photon number basis as $\hat \rho_\text{in} = \sum_{m,n} \rho_\text{in}^{mn} \ket{m}\!\bra{n}$,
the corresponding Wigner function can also be expanded as
$W_\text{in}(\xi) = \sum_{m,n} \rho_\text{in}^{mn} W^{\ket{m}\!\bra{n}}(\xi)$.
By using Eq.~(\ref{imperfect_EPR_GC}) and linearity of convolution,
the Wigner function of the teleported state $\hat{\rho}_\text{out}$ can be written as
\begin{align}
W_\text{out} (\xi )
= \sum_{m,n} \rho_\text{in}^{mn} W^{\ket{m}\!\bra{n}}_\text{out} (\xi ),
\label{outuput_wigner_photon_basis_decomp}
\end{align}
where we define
\begin{align}
W^{\hat{A}}_\text{out} (\xi ) &\equiv \frac{1}{g^2} \left[W^{\hat{A}} \circ G_{\tau} \right] \left( \frac{\xi}{g} \right).
\label{eq:teleported_A}
\end{align}
The Wigner functions in Eq.~(\ref{outuput_wigner_photon_basis_decomp}) can be converted to their corresponding density matrices.
From $W_\text{out} (\xi )$, each density matrix element $\braket{j|\hat{\rho}_\text{out}|k}$
is obtained by~\cite{01Wolf}
\begin{align}
&\braket{j|\hat \rho_\text{out}|k}= 2\pi \iint d\xi W_\text{out}( \xi ) W^{\ket{k}\!\bra{j}} (\xi ) \nonumber\\
&\quad=\sum_{m,n}\rho_\text{in}^{mn}\cdot 2\pi\iint d\xi W^{\ket{m}\!\bra{n}}_\text{out}( \xi ) W^{\ket{k}\!\bra{j}} (\xi ).
\label{eq:element_after_teleportation}
\end{align}
Next we define a transition operator $\hat{T}(\hat{A})$ as
\begin{align}
\hat{T}(\hat{A})=\sum_{j,k}2\pi \iint d\xi W_\text{out}^{\hat{A}}( \xi ) W^{\ket{k}\!\bra{j}} (\xi )\ket{j}\!\bra{k},
\label{eq:transition_operator}
\end{align}
describing a map from an operator to an operator: $\hat{A}\mapsto \hat{T}(\hat{A})$.
This map describes the transition of a density matrix element $\hat{A}$ through the CV teleportation channel of Eq.~(\ref{imperfect_EPR_GC}).
By substituting Eq.~(\ref{eq:transition_operator}) into Eq.~(\ref{eq:element_after_teleportation}), we get
\begin{align}
\braket{j|\hat \rho_\text{out}|k}&=\braket{j|\sum_{m,n}\rho_\text{in}^{mn}\hat{T}(\ket{m}\!\bra{n})|k} \nonumber\\
\Longleftrightarrow\quad
\hat \rho_\text{out}&= \sum_{m,n} \rho_\text{in}^{mn} \hat{T}\bigl({\ket{m}\!\bra{n}}\bigr).
\label{outuput_density_decomp}
\end{align}
Thus, the output density matrix is obtained by replacing $\ket{m}\!\bra{n}\to\hat{T}\bigl({\ket{m}\!\bra{n}}\bigr)$
for any given input density matrix $\hat \rho_\text{in} = \sum_{m,n} \rho_\text{in}^{mn} \ket{m}\!\bra{n}$.
In this picture,
the final density matrix at the output is a combination of
each density matrix element teleported independently.

Before applying this transition-operator formalism to qubits,
we derive the expression of
$\hat{T}\bigl({\ket{m}\!\bra{n}}\bigr)=\sum_{j,k}T_{mn\to jk}\ket{j}\!\bra{k}$ for $m,n= 0,1$, where the coefficient
\begin{align}
T_{mn\to jk}\equiv2\pi \iint d\xi W_\text{out}^{\ket{m}\!\bra{n}}( \xi ) W^{\ket{k}\!\bra{j}} (\xi )
\label{eq:transition_coef}
\end{align}
represents the transition probability for a component $\ket{m}\!\bra{n}$ to be transformed into $\ket{j}\!\bra{k}$ by teleportation.
To calculate this, we obtain the following functions from Eq.~(\ref{photon_num_wigner}) and (\ref{eq:teleported_A}),
\begin{align}
W^{\ket{0}\bra{0}}_\text{out}(\xi )\!&=\!\frac{1}{\lambda\pi} \exp \left[-\xi^T\xi/\lambda\right], \label{wigner_00} \\
W^{\ket{1}\bra{0}}_\text{out}(\xi )\!&=\!W^{\ket{0}\bra{1}}_\text{out}(\sigma_z\xi )\!=\!\frac{\sqrt{2}  v^T\xi g}{\lambda^2 \pi}  \exp\!\left[-\xi^T\xi/\lambda \right], \label{wigner_01} \\
W^{\ket{1}\bra{1}}_\text{out}(\xi )\!&=\!\frac{\left( 2g^2\xi^T \xi+\lambda(\lambda-2g^2)\right)}{\lambda^3 \pi} \exp\!\left[-\xi^T\xi/\lambda \right]. \label{wigner_11}
\end{align}
Here we introduced a new parameter $\lambda\equiv g^2(2\tau +1)$ for simplicity.
The condition $\lambda\ge1$ can be proven from Eq.~(\ref{gamma}), and the equality is attained if and only if $l=0$ and $g=\tanh r$.
By substituting the above functions and Eq.~(\ref{photon_num_wigner}) into Eq.~(\ref{eq:transition_coef}), we obtain
\begin{align}
T_{00\to jk}\!&=\!\frac{2(\lambda-1)^k}{(\lambda+1)^{k+1}} \delta_{j,k}, \label{photon_dis_00} \\
T_{10\to jk}\!&=\!T_{01\to kj}
\!=\!\frac{4g\sqrt{k+1}(\lambda-1)^k}{(\lambda +1)^{k+2}}\delta_{j,k+1}, \label{photon_dis_10} \\
T_{11\to jk}\!&=\!\frac{2(\lambda\!-\!1)^{k\!-\!1}}{(\lambda\!+\!1)^{k\!+\!2}} \!
\left[(\lambda\!-\!2g^2\!+\!1)\!(\lambda\!-\!1)\!+\!4kg^2\right]\! \delta_{j,k}, \label{photon_dis_11}
\end{align}
where $\delta_{j,k}$ is the discrete Kronecker delta function.
All the coefficients above are non-negative.
It can be seen that the diagonal elements $\ket{0}\!\bra{0}$ and $\ket{1}\!\bra{1}$ (off-diagonal elements $\ket{1}\!\bra{0}$ and $\ket{0}\!\bra{1}$)
of the input state contribute only to the diagonal elements $\ket{k}\!\bra{k}$
(off-diagonal elements $\ket{k+1}\!\bra{k}$ and $\ket{k}\!\bra{k+1}$) in the output state.

The formulas derived above enable us to describe CV teleportation of qubits fully
in terms of density matrices.
For the single-rail qubit in Eq.~(\ref{single_rail_qubit}), the teleportation process is then expressed by
\begin{align}
&\ket{\psi}_\text{s}\!\bra{\psi}
\!=\!|\alpha|^2\!\ket0\!\bra0\!+\!\alpha^*\!\beta\!\ket1\!\bra0
\!+\!\alpha\beta^*\!\ket0\!\bra1\!+\!|\beta|^2\!\ket1\!\bra1 \nonumber\\
&\Rightarrow
\hat{T}(\ket{\psi}_\text{s}\!\bra{\psi})\!=\!
|\alpha|^2\hat{T}^{00}\!+\!\alpha^*\beta\hat{T}^{10}
\!+\!\alpha\beta^*\hat{T}^{01}\!+\!|\beta|^2\hat{T}^{11}\!,
\label{eq:CVQT_single_rail}
\end{align}
defining $\hat{T}^{mn}\equiv\hat{T}(\ket{m}\!\bra{n})$.
In the case of the dual-rail qubit in Eq.~(\ref{dual_rail_qubit}),
each mode is transmitted through a teleportation channel independently.
Thus, we extend the single-mode transition operator to a two-mode transition operator, $\hat{T}_\text{XY}(\hat\rho_\text{XY})$,
for any two-mode state $\hat{\rho}_\text{XY}=\sum_{j,k,m,n}\rho_{jkmn}\ket{j}_\text{X}\!\bra{k}\otimes\ket{m}_\text{Y}\!\bra{n}$ as
\begin{align}
\hat{T}_\text{XY}(\hat{\rho}_\text{XY})\!\equiv\!
\sum_{j,k,m,n}\!\rho_{jkmn}\hat{T}_\text{X}(\ket{j}_\text{X}\!\bra{k})\!\otimes\!\hat{T}_\text{Y}(\ket{m}_\text{Y}\!\bra{n}).
\end{align}
In general, the two teleportation channels $\hat{T}_\text{X}$ and $\hat{T}_\text{Y}$ are characterized by different parameters $(\tau, g)$ according to Eq.~(\ref{imperfect_EPR_GC}).
Thus, teleportation of a dual-rail qubit can be described by
\begin{align}
&\ket{\psi}_\text{d}\!\bra{\psi}
=|\alpha|^2\ket0_\text{X}\!\bra0\!\otimes\!\ket1_\text{Y}\!\bra1
+\alpha^*\beta\ket1_\text{X}\!\bra0\!\otimes\!\ket0_\text{Y}\!\bra1 \nonumber\\
&\quad\quad\quad\quad+\alpha\beta^*\ket0_\text{X}\!\bra1\!\otimes\!\ket1_\text{Y}\!\bra0
+|\beta|^2\ket1_\text{X}\!\bra1\!\otimes\!\ket0_\text{Y}\!\bra0 \nonumber\\
&\Rightarrow
\hat{T}_\text{XY}(\ket{\psi}_\text{d}\!\bra{\psi})\!=\!
|\alpha|^2\hat{T}_\text{X}^{00}\!\otimes\!\hat{T}_\text{Y}^{11}
+\alpha^*\beta\hat{T}_\text{X}^{10}\!\otimes\!\hat{T}_\text{Y}^{01} \nonumber\\
&\hspace{25mm}+\alpha\beta^*\hat{T}_\text{X}^{01}\!\otimes\!\hat{T}_\text{Y}^{10}
+|\beta|^2\hat{T}^{11}_\text{X}\!\otimes\!\hat{T}_\text{Y}^{00},
\label{eq:CVQT_dual_rail}
\end{align}
where $\hat{T}_{i}^{mn}\equiv\hat{T}_{i}(\ket{m}_i\!\bra{n})$ ($i=$X, Y).
Equations~(\ref{eq:CVQT_single_rail}) and (\ref{eq:CVQT_dual_rail}),
together with the coefficients given by Eqs.~(\ref{photon_dis_00})-(\ref{photon_dis_11}),
give the output density matrices as functions of
squeezing parameter $r$, loss $l$, and gain $g$;
and hence gain tuning for teleporting DV states can be investigated
in detail from these results.

One advantage of our formalism compared to the transfer-operator formalism~\cite{01Ide,02Ide}
is that an actual, realistic experiment can be more accurately modeled by taking into account the impurity of squeezing.
In addition, our formalism is applicable to more complex channels than a simple teleportation channel
by replacing the parameters $(\tau,g)$ for each $\hat{T}$ by $(\tau_\text{tot},g_\text{tot})$
[namely, Eq.~(\ref{imperfect_EPR_GC}) is replaced by Eq.~(\ref{eq:overall_channel})].
It can also be straightforwardly extended to the multi-qubit situation where all or some parts of the qubits are teleported.
Possible applications are various hybrid teleportation experiments,
such as CV teleportation of a multi-qubit system,
sequential CV teleportation of qubits,
and entanglement swapping~\cite{93Zukowki} using a DV entangled state and CV teleportation.

%%%%%%%%%%%%%%%%% DR-qubit teleportation %%%%%%%%%%%%%%%%%%%%%%%%%%%%%%%%%%%%%%%%%%%%%%%%%%%%%%%%%%%%%%%%%%%%%

\section{Teleportation of dual-rail qubits}\label{sec:qubit_tele}

Continuous-variable teleportation of a dual-rail qubit is one of the most important and fundamental examples of optical hybrid quantum information processing,
as was experimentally demonstrated recently~\cite{13Takeda}.
The ideal situation of this experiment has been modeled already by the transfer-operator formalism
on the assumption that the input qubit state and the resource squeezed states are perfectly pure~\cite{01Ide,02Ide}.
This assumption is, however, not true in the actual experiment.
Here we model this teleportation experiment more accurately by taking into account the impurity of the input state as well as the squeezed states
(all measurement inefficiencies can be incorporated into the initial losses as shown in Appendix B).
Since fidelity is used as a figure of merit in Ref.~\cite{13Takeda},
we investigate the optimal gain to obtain the maximum fidelity under various experimental conditions.
The theoretical results here are shown to be in good agreement with the experimental results in Ref.~\cite{13Takeda}.

We start with describing the output density matrix of teleportation by using transition operators.
The experimental dual-rail qubit input can be modeled by
a mixed state of the pure qubit in Eq.~(\ref{dual_rail_qubit}) and a two-mode vacuum state as
\begin{align}
\hat \rho_\text{in} &=
\eta \ket{\psi}_\text{d}\!\bra{\psi}
+(1-\eta ) \ket{0}_\text{X}\!\bra{0}\otimes\ket{0}_\text{Y}\!\bra{0},
\label{input_density}
\end{align}
where $\eta$ is the fraction of the qubit ($0\leq \eta \leq 1$).
As is mentioned in Sec.~\ref{sec:vac_single},
CV teleportation of this qubit requires two parallel teleportation channels, which have different parameters $(\tau,g)$ in general.
% In this case the output density matrix is calculated by $\hat{\rho}_\text{out}=\hat{T}_\text{XY}(\hat\rho_\text{in})$ of Eq.~(\ref{eq:CVQT_dual_rail}).
However, we may assume the same $(\tau,g)$ for the two teleportation channels
when both rails of the qubit~\cite{12Takeda} are teleported by the same CV teleporter, as in Ref.~\cite{13Takeda}.
This assumption greatly simplifies the description of the experiment in the following two aspects.
Firstly, for an arbitrary qubit $\ket{\psi}_\text{d}$,
its density matrix $\hat{\rho}_\text{in}$ in Eq.~(\ref{input_density}) can be decomposed into a tensor product
\begin{align}
\hat{U}\hat\rho_\text{in}\hat{U}^\dagger=
\left[\eta\ket{1}_\text{X}\!\bra{1}+(1-\eta)\ket{0}_\text{X}\!\bra{0}\right]
\otimes \ket{0}_\text{Y}\!\bra{0}
\label{eq:decomposed_input}
\end{align}
via a beam-splitter transformation defined by a unitary operator $\hat{U}$ satisfying
$\hat U\hat a_\text{X}^\dagger \hat U^\dagger  = \beta^* \hat a_\text{X}^\dagger - \alpha \hat a_\text{Y}^\dagger$
and $\hat U\hat a_\text{Y}^\dagger \hat U^\dagger  = \alpha^* \hat a_\text{X}^\dagger + \beta \hat a_\text{Y}^\dagger$
($\hat{a}^\dagger_i$ denotes a creation operator of mode $i$).
Secondly, when the teleportation channels for both rails have the same $(\tau,g)$,
the basis transformation $\hat\rho_\text{XY}\to\hat{U}\hat{\rho}_\text{XY}\hat{U}^\dagger$ and
the teleportation map $\hat\rho_\text{XY}\to\hat{T}_\text{XY}(\hat\rho_\text{XY})$ of Eq.~(\ref{eq:CVQT_dual_rail})
commute for any two-mode state $\hat{\rho}_\text{XY}$:
\begin{equation}
\hat{T}_\text{XY}\left(\hat{U}\hat{\rho}_\text{XY}\hat{U}^\dagger\right)=
\hat{U}\left[ \hat{T}_\text{XY}\left(\hat{\rho}_\text{XY}\right)\right]\hat{U}^\dagger.
\label{eq:commutation_TU}
\end{equation}
This is proven in Appendix A.
These two properties together mean
that the density matrix of the teleported state $\hat{\rho}_\text{out}=\hat{T}_\text{XY}(\hat\rho_\text{in})$ satisfies
\begin{align}
\hat{U}\hat \rho_\text{out}\hat{U}^\dagger
\!=\!\hat{T}_\text{XY}\!\left(\!\hat{U}\hat{\rho}_\text{in}\hat{U}^\dagger\!\right)
\!=\!\left[\eta\hat{T}_\text{X}^{11}
\!+\!(1\!-\!\eta)\hat{T}_\text{X}^{00}\right]
\!\otimes\!\hat{T}_\text{Y}^{00}.
\label{eq:decomposed_output}
\end{align}
Equations (\ref{eq:decomposed_input}) and (\ref{eq:decomposed_output})
indicate that, after the transformation $\hat{U}$,
dual-rail qubit teleportation can be interpreted as parallel teleportation
of a single photon with loss (mode X) and a vacuum (mode Y).
From Eqs.~(\ref{photon_dis_00}) and (\ref{photon_dis_11}),
it is shown that $\hat{U}\hat \rho_\text{out}\hat{U}^\dagger$
in Eq.~(\ref{eq:decomposed_output}) has only diagonal density matrix elements.
The matrix elements of $\hat \rho_\text{out}$ can be obtained by the inverse transformation $\hat{U}^\dagger$ of Eq.~(\ref{eq:decomposed_output}).
Since $\hat{U}^\dagger$ preserves the total photon number,
i.e., ${}_\text{X}\!\bra{j}{}_\text{Y}\!\bra{k}\hat{U}^\dagger \ket{m}_\text{X}\ket{n}_\text{Y}=0$ for $j+k \neq m+n$,
$\hat{\rho}_\text{out}$ has non-zero $\ket{j}_\text{X}\ket{k}_\text{Y}\bra{m}_\text{X}\bra{n}_\text{Y}$ elements only when $j+k=m+n$.
As a result, the element of $\hat{\rho}_\text{out}$ belongs to either
the vacuum subspace spanned by $\{\ket{0}_\text{X}\ket{0}_\text{Y}\}$,
or the single-photon subspace spanned by $\{\ket{0}_\text{X}\ket{1}_\text{Y},\ket{1}_\text{X}\ket{0}_\text{Y}\}$,
or the two-photon subspace spanned by $\{\ket{0}_\text{X}\ket{2}_\text{Y},\ket{1}_\text{X}\ket{1}_\text{Y},\ket{2}_\text{X}\ket{0}_\text{Y}\}$,
and so on. Here, the single-photon subspace is the original qubit subspace where the quantum information is encoded.

Now the fidelity between $\hat{\rho}_\text{in}$ and $\hat{\rho}_\text{out}$ can be calculated to assess the performance of teleportation.
Two types of fidelity are introduced in Ref.~\cite{13Takeda}.
One is the overall transfer fidelity $F_\text{state}$
which is directly calculated from $\hat \rho _\text{in}$ and $\hat \rho _\text{out}$ as~\cite{94Joz}
\begin{align}
F_\text{state}=\left[\text{Tr}\left(\sqrt{\sqrt{\hat{\rho}_\text{in}}\hat{\rho}_\text{out}\sqrt{\hat{\rho}_\text{in}}}\right)\right]^2.
\label{fidelity}
\end{align}
This fidelity reflects the entire two-mode Hilbert space, taking into account the vacuum and multi-photon contributions.
Thus, $F_\text{state}$ describes the performance of a ``deterministic'' teleportation
that does not pre-select or post-select specific parts of the quantum states.
Since the fidelity between $\hat{\rho}_\text{in}$ and $\hat{\rho}_\text{out}$ is equal to
the fidelity between $\hat{U}\hat{\rho}_\text{in}\hat{U}^\dagger$ and $\hat{U}\hat{\rho}_\text{out}\hat{U}^\dagger$~\cite{94Joz},
it is straightforwardly calculated from Eqs.~(\ref{eq:decomposed_input}) and (\ref{eq:decomposed_output}) as
\begin{align}
F_\text{state}=&
T_{00\to00}\!\left[ \sqrt{\eta\left(\eta T_{11\to11}\!+\!(1-\eta)T_{00\to11}\right)}\right.\nonumber\\
&\quad\left.+\sqrt{(1\!-\!\eta)\!\left(\eta T_{11\to00}\!+\!(1\!-\!\eta)T_{00\to00}\right) }\right]^2,
\label{eq:Fstate}
\end{align}
with coefficients $T$ given by Eqs.(\ref{photon_dis_00}) and (\ref{photon_dis_11}).
The other fidelity in Ref.~\cite{13Takeda} is
an indicator to assess the qubit components alone:
\begin{align}
F_\text{qubit}={}_\text{d}\!\bra{\psi}\hat\rho^\text{qubit}_\text{out}\ket{\psi}_\text{d},
\label{F_qubit}
\end{align}
where $\ket{\psi}_\text{d}$ is the pure input qubit in Eq.~(\ref{dual_rail_qubit})
and $\hat \rho^\text{qubit}_\text{out}$ is the matrix obtained by extracting and renormalizing the qubit subspace
spanned by $\{ \ket{0}_\text{X}\ket{1}_\text{Y}, \ket{1}_\text{X}\ket{0}_\text{Y}\}$.
$F_\text{qubit}$ can be calculated as the fidelity between
$\hat{U}\ket{\psi}_\text{d}\!\bra{\psi}\hat{U}^\dagger=\ket{1}_\text{X}\!\bra{1}\otimes\ket{0}_\text{Y}\!\bra{0}$
and $\hat{U}\hat{\rho}_\text{out}^\text{qubit}\hat{U}^\dagger$.
The latter is obtained by extracting the terms with one photon in total
from $\hat{U}\hat{\rho}_\text{out}\hat{U}^\dagger$ and then renormalizing.
Thus we obtain
\begin{align}
F_\text{qubit}=\frac{{P}_\text{trans}}{{P}_\text{trans}+{{P}_\text{flip}}},
\label{eq:Fqubit}
\end{align}
where
\begin{align}
{P}_\text{trans}&=T_{00\to00}\left[\eta T_{11\to11}+(1-\eta)T_{00\to11}\right]\\
{P}_\text{flip}&=T_{00\to11}\left[\eta T_{11\to00}+(1-\eta)T_{00\to00}\right] \label{eq:Pflip}
\end{align}
are the probabilities that
the photon number is transferred correctly
($\ket{1}_\text{X}\ket{0}_\text{Y}\to\ket{1}_\text{X}\ket{0}_\text{Y}$)
or flipped ($\ket{1}_\text{X}\ket{0}_\text{Y}\to\ket{0}_\text{X}\ket{1}_\text{Y}$).
The sum $P_\text{qubit}={P}_\text{trans}+{P}_\text{flip}$ gives the probability of obtaining the qubit at the output.
Note that both fidelities, $F_\text{state}$ in Eq.~(\ref{eq:Fstate}) and $F_\text{qubit}$ in Eq.~(\ref{eq:Fqubit}),
are independent of the qubit coefficients $\alpha$ and $\beta$ in Eq.~(\ref{dual_rail_qubit}).
This indicates that the optimal gain is also independent of the initial qubit state.

\begin{figure*}[!tb]
\centering
\subfigure[]{
\hspace{1mm}\includegraphics[clip,scale=0.3]{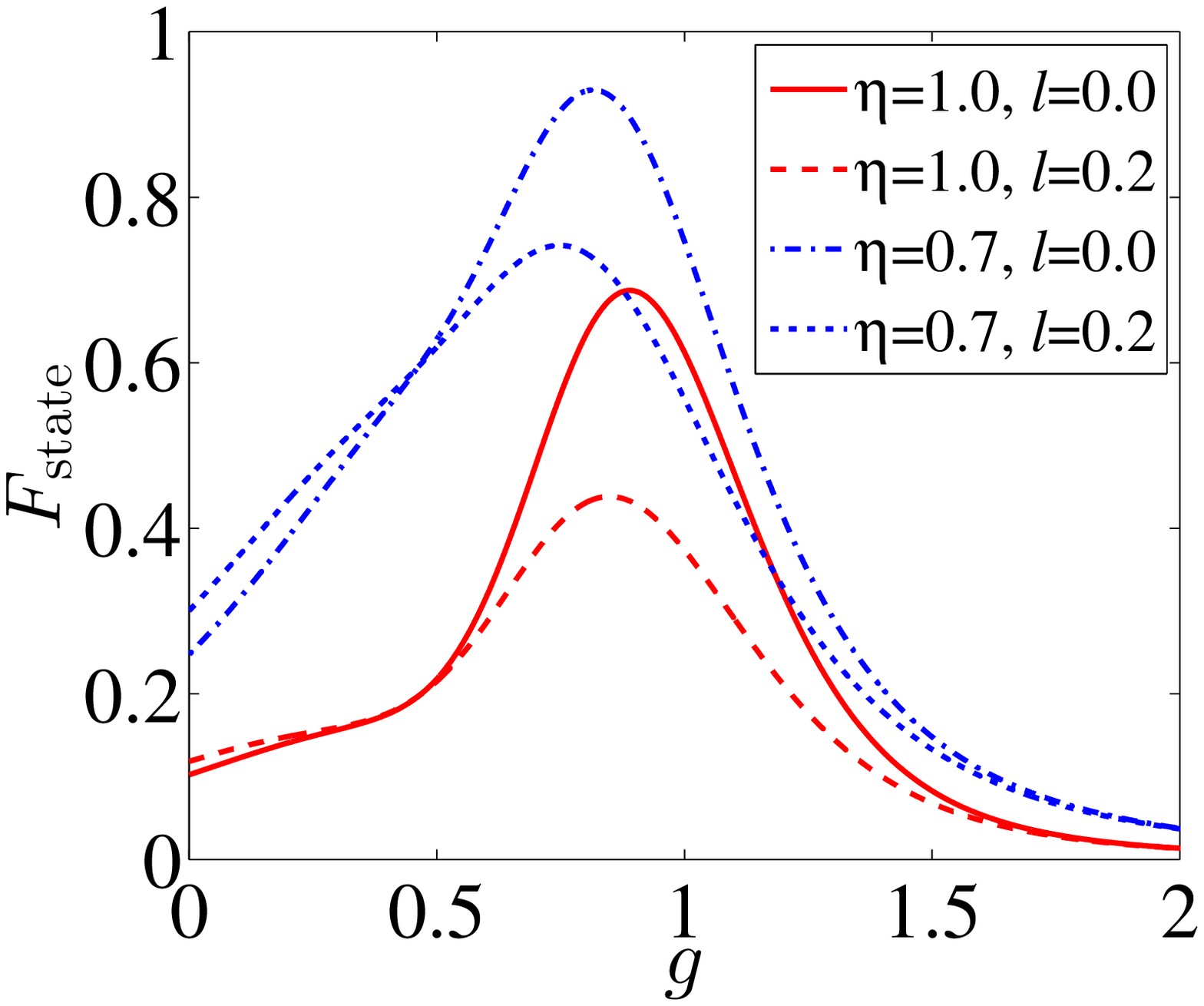}\hspace{1mm}
\label{sfig:F_state_gain_ps}
}
\subfigure[]{
\hspace{1mm}\includegraphics[clip,scale=0.3]{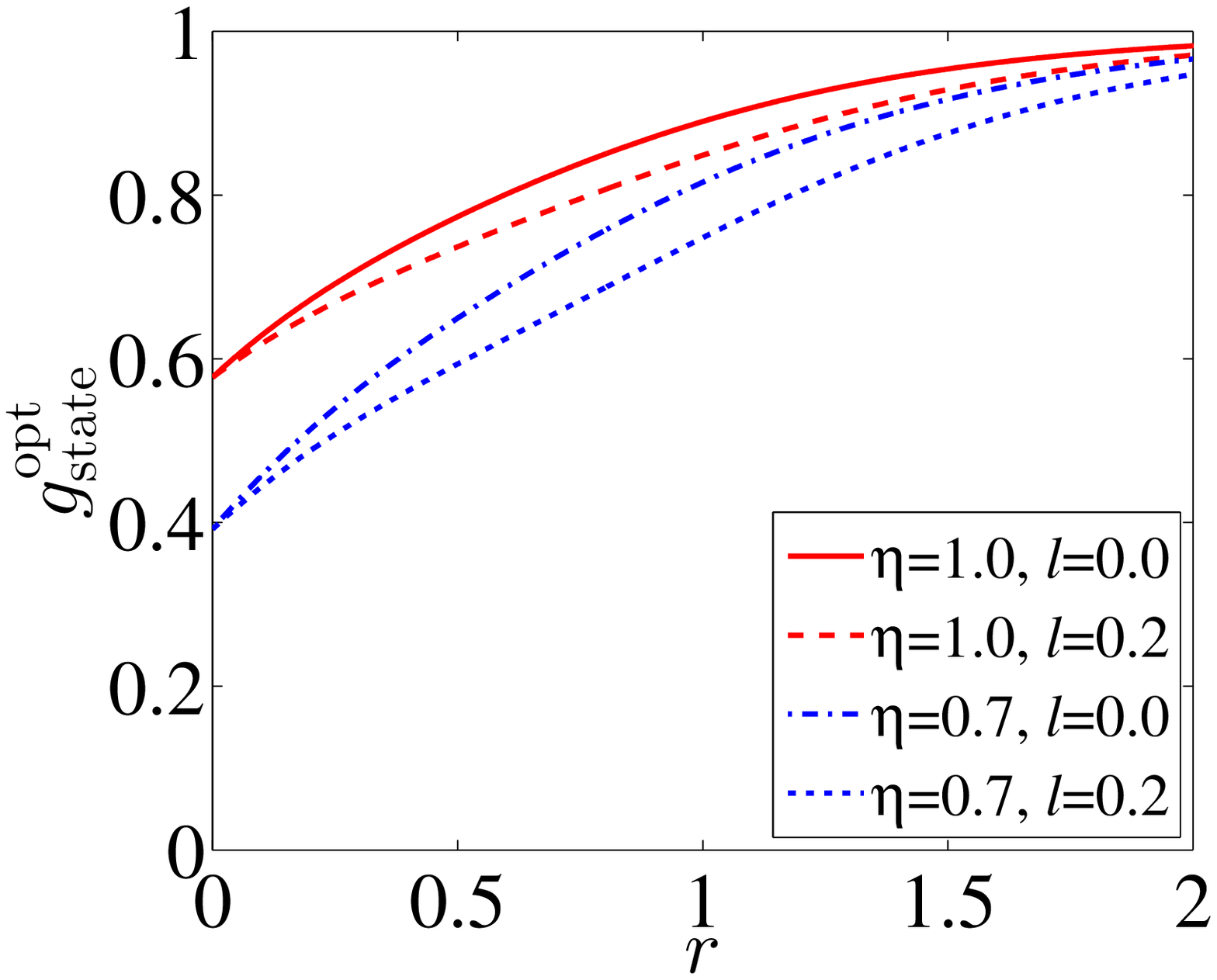}\hspace{1mm}
\label{sfig:F_state_g_opt_ps}
}
\subfigure[]{
\hspace{1mm}\includegraphics[clip,scale=0.3]{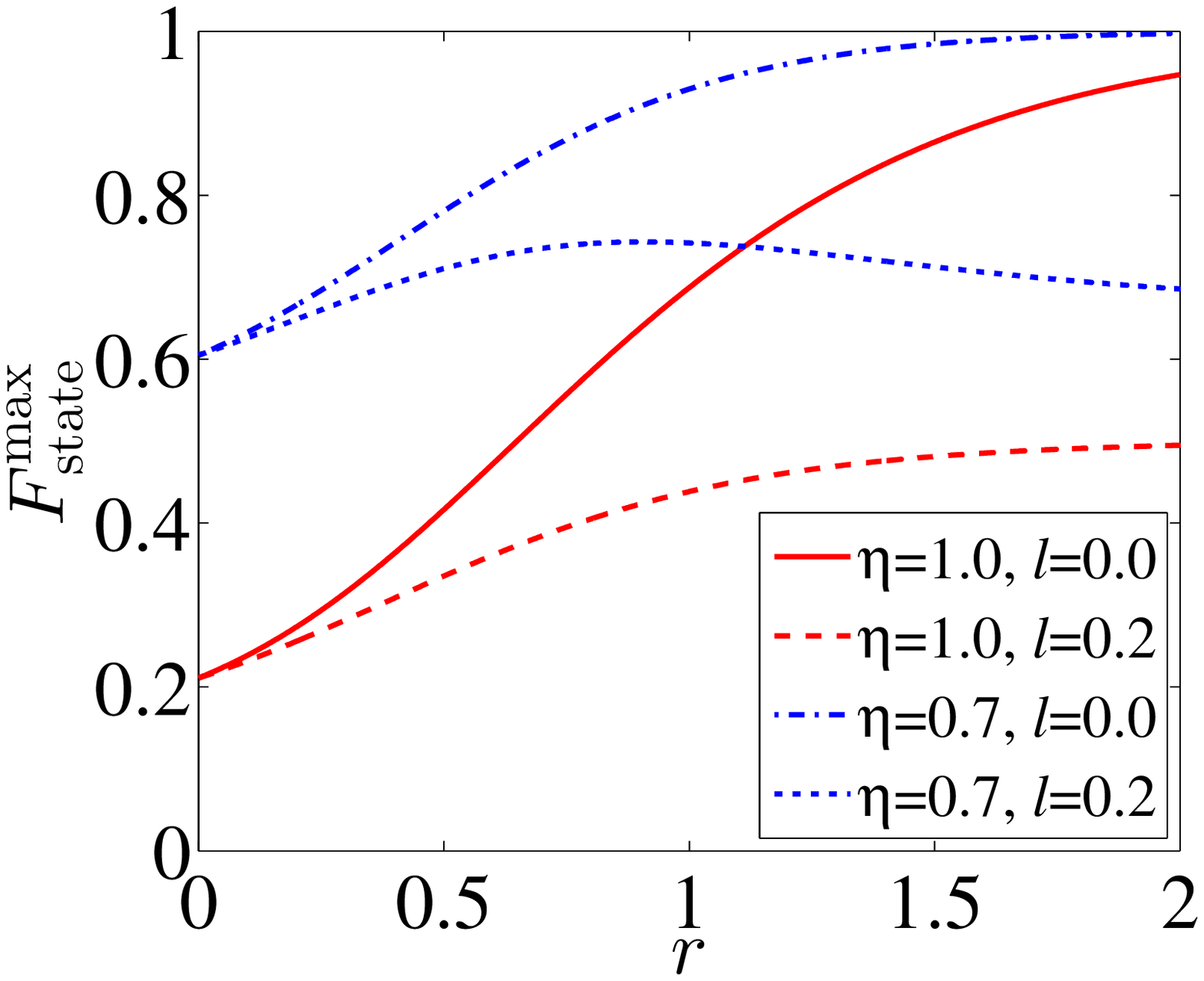}\hspace{1mm}
\label{sfig:F_state_g_opt_fidelity_ps}
}
\subfigure[]{
\hspace{1mm}\includegraphics[clip,scale=0.3]{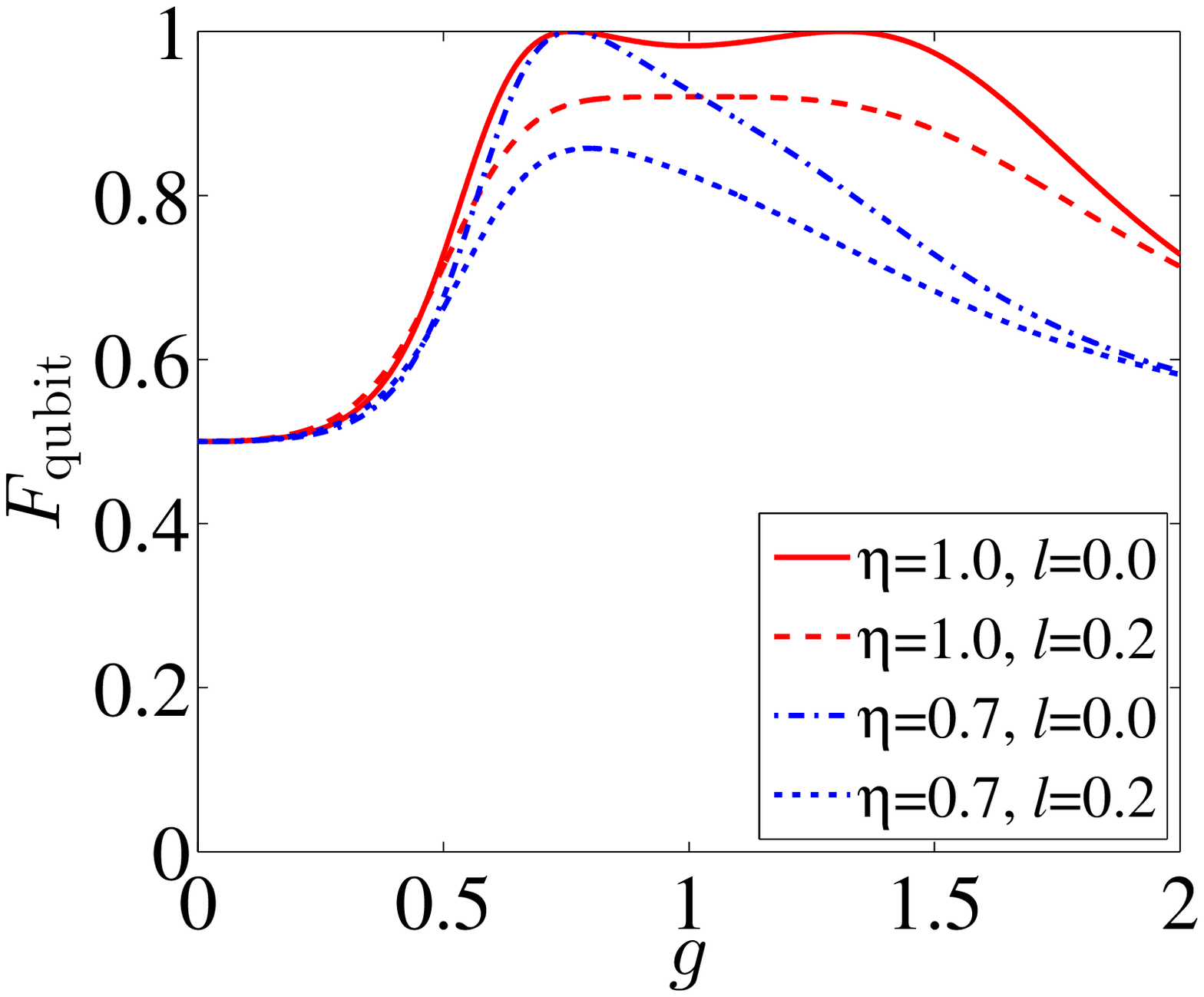}\hspace{1mm}
\label{sfig:F_qubit_gain_ps}
}
\subfigure[]{
\hspace{1mm}\includegraphics[clip,scale=0.3]{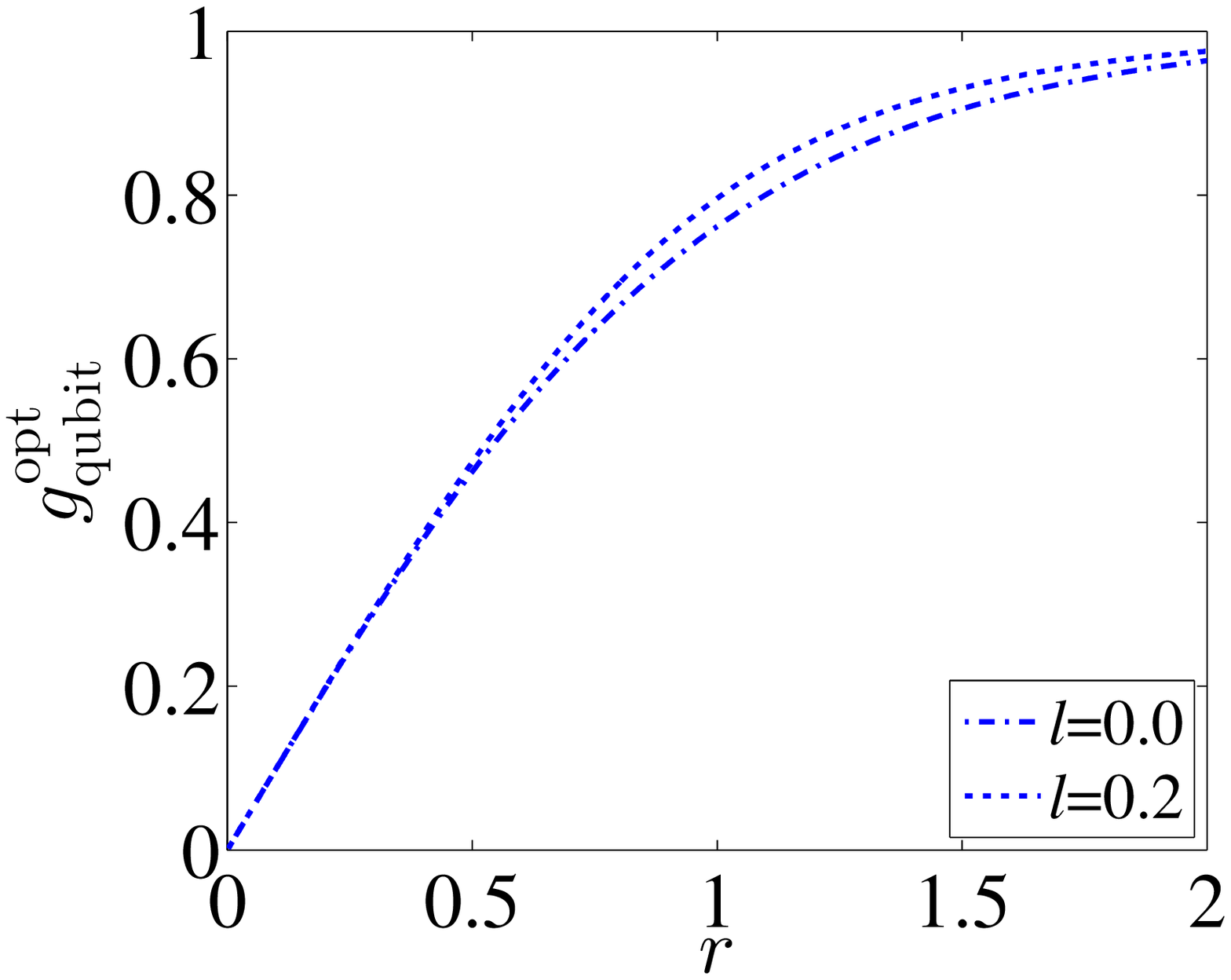}\hspace{1mm}
\label{sfig:F_qubit_g_opt_ps}
}
\subfigure[]{
\hspace{1mm}\includegraphics[clip,scale=0.3]{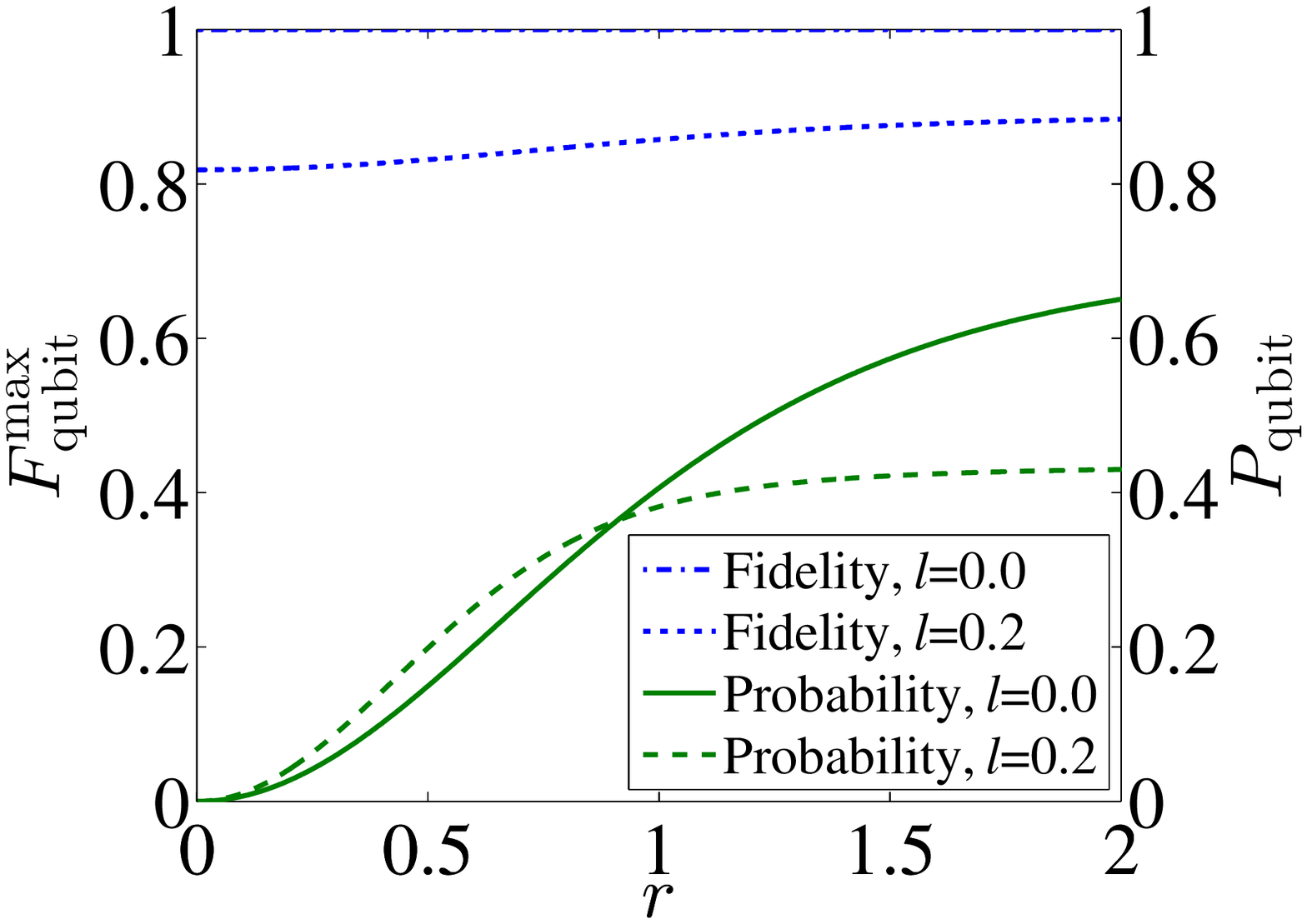}\hspace{1mm}
\label{sfig:F_qubit_g_opt_fidelity_ps}
}
\subfigure[]{
\hspace{1mm}\includegraphics[clip,scale=1.2]{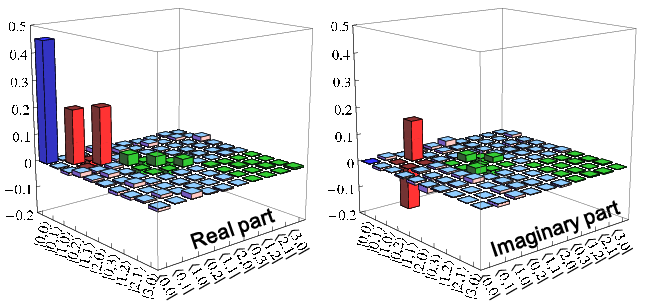}\hspace{1mm}
\label{sfig:experiment_result_tele}
}
\subfigure[]{
\hspace{1mm}\includegraphics[clip,scale=1.2]{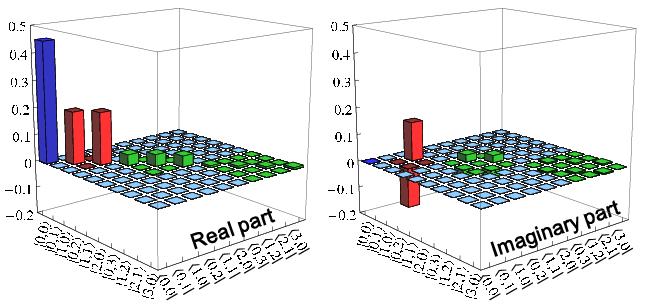}\hspace{1mm}
\label{sfig:simulation_result_tele}
}
\caption{(color online) Simulation results of CV teleportation of a dual-rail qubit.
(a) $g$ dependence of $F_\text{state}$ at $r=1.0$.
(b) $r$ dependence of $g_\text{state}^\text{opt}$.
(c) $r$ dependence of $F_\text{state}^\text{max}$.
(d) $g$ dependence of $F_\text{qubit}$ at $r=1.0$.
(e) $r$ dependence of $g_\text{qubit}^\text{opt}$ at $\eta=0.7$.
(f) $r$ dependence of $F_\text{qubit}^\text{max}$ and $P_\text{qubit}$ at $\eta=0.7$.
(g) Experimental $\hat{\rho}_\text{out}$ in Ref.~\cite{12Takeda}.
(h) Theoretical $\hat{\rho}_\text{out}$ in Eq.~(\ref{eq:decomposed_output})
for $(\alpha,\beta,\eta_1,\eta_2,r,l,g)=(1/\sqrt2,-i/\sqrt2,0.69,0.06,1.01,0.2,0.79)$.}
\label{fig:tele_simulation_results}
\end{figure*}

Now we discuss the dependence of fidelities $F_\text{state}$ and $F_\text{qubit}$ on the four experimental parameters $(\eta,r,l,g)$
using Eqs.~(\ref{eq:Fstate}) and (\ref{eq:Fqubit}).
Here $1-\eta$ and $l$ are experimental losses, constant for each experimental setup.
Therefore only the $g$ and $r$ dependences of the fidelities are plotted for fixed $(\eta,l)$ values
(e.g. in Ref.~\cite{13Takeda}, $\eta\sim0.7$ and $l\sim0.2$ are given,
while $r\in\{0.71,1.01,1.56\}$ and $g\in\{0.5,0.63,0.79,1.0\}$ are varied).
First the $g$ dependence of $F_\text{state}$ at $r=1.0$ is plotted for various $(\eta,l)$ values in Fig.~\ref{sfig:F_state_gain_ps}.
Here the parameters $(\eta,r,l)=(0.7,1.0,0.2)$ reflect one of the experimental settings in Ref.~\cite{13Takeda}.
It can be seen that gain tuning greatly improves $F_\text{state}$ from the unit-gain regime at a certain gain below unity.
However, the optimal gain $g^\text{opt}_\text{state}$ to give the maximum $F_\text{state}$ differs according to $(\eta,l)$.
Note that generally $F_\text{state}$ is higher for more mixed input states, leading to a higher 
$F_\text{state}$ for $\eta=0.7$ than $\eta=1.0$.
The $r$ dependence of $g^\text{opt}_\text{state}$ and the maximum fidelity $F^\text{max}_\text{state}$ at this gain
is shown in Figs.~\ref{sfig:F_state_g_opt_ps} and \ref{sfig:F_state_g_opt_fidelity_ps} for various $(\eta,l)$ values.
For $l=0$, we can see $F_\text{state}^\text{max}\to1$ in the limit of $r\to\infty$,
but $F_\text{state}^\text{max}$ does not reach unity for $l>0$.
More interestingly, in the case of $l>0$ and $\eta<1$, there is an optimal squeezing level $r^\text{opt}_\text{state}$ which gives the maximum $F_\text{state}^\text{max}$
[\textit{e.g.}, $r^\text{opt}_\text{state}=0.91$ for $(\eta,l)=(0.7,0.2)$].
This result agrees with the fact that
the highest value of $F_\text{state}^\text{max}$ was obtained at $r=1.01$ among $\{0.71,1.01,1.56\}$
in the experiment of Ref.~\cite{13Takeda}.

Next we examine $F_\text{qubit}$ in the same way.
Figure~\ref{sfig:F_qubit_gain_ps} shows the $g$ dependence of $F_\text{qubit}$ at $r=1.0$ for various $(\eta,l)$ values.
It is notable that $F_\text{qubit}$ reaches unity for certain conditions even though the squeezing parameter $r$ is finite.
One condition is $g=g_\text{att}(=0.76)$ and $\eta>0$, when the teleportation is equivalent to a pure attenuation channel.
Here the teleportation only increases the vacuum component of the initial qubit state of Eq.~(\ref{input_density}),
which leads to $T_{00\to11}=0$ and thus $P_\text{flip}=0$ in Eq.~(\ref{eq:Pflip}).
The other condition is $g=g_\text{amp}(=1.31)$ and $\eta=1$,
when the initial pure qubit is amplified and transformed into the mixture of a qubit and more-than-one photon states.
This condition also gives $P_\text{flip}=0$ since $T_{11\to00}=0$ and $1-\eta=0$ in Eq.~(\ref{eq:Pflip}).
In most of the realistic conditions of $\eta<1$ and $l>0$, the peak at $g=g_\text{amp}$ vanishes and $F_\text{qubit}$ is maximal when the gain is close to $g_\text{att}$.
For $l>0$, the maximum $F_\text{qubit}$ is limited below unity due to the inevitable thermalization effect of teleportation, as mentioned in Sec.~\ref{sec:vac_single}.
Thus, the loss on the resource EPR state limits the experimental $F_\text{qubit}$ to at best 0.90 in Ref.~\cite{13Takeda}.
The $r$ dependence of the optimal gain $g^\text{opt}_\text{qubit}$,
the maximum fidelity $F^\text{max}_\text{qubit}$, and the probability $P_\text{qubit}$ at this gain
are plotted in Figs.~\ref{sfig:F_qubit_g_opt_ps} and \ref{sfig:F_qubit_g_opt_fidelity_ps} (we fixed $\eta=0.70$).
In Fig.~\ref{sfig:F_qubit_g_opt_ps}, $g_\text{qubit}^\text{opt}=g_\text{att}$ for $l=0$ (for any $0<\eta<1$).
The finite loss of $l=0.2$ leads to only a small discrepancy between $g_\text{qubit}^\text{opt}$ and $\tanh r$;
for $l=0.2$, $g_\text{qubit}^\text{opt}$ is closer to $\tanh r$ for smaller $\eta$.
Thus, in most cases, choosing $g=\tanh r$ regardless of $(\eta,l)$ is nearly optimal for faithful qubit information transfer.
Figure~\ref{sfig:F_qubit_g_opt_fidelity_ps} shows that
$P_\text{qubit}$ increases with increasing $r$, but the increment of $F^\text{max}_\text{qubit}$ is small.
$F^\text{max}_\text{qubit}$ is mainly limited by $l$ (not by $r$)
which defines the minimal thermalization effect of teleportation.

Finally,  we directly compare the theoretical $\hat{\rho}_\text{out}$ in Eq.~(\ref{eq:decomposed_output})
and the experimental $\hat{\rho}_\text{out}$ in Ref.~\cite{13Takeda} for the purpose of demonstrating the validity of our model.
One of the experimental results along with its theoretically calculated simulation result for the same parameters $(\alpha,\beta,\eta,r,l,g)$
is illustrated in Figs.~\ref{sfig:experiment_result_tele} and \ref{sfig:simulation_result_tele};
they are in good agreement.
The fidelities between theoretical and experimental $\hat{\rho}_\text{out}$
for 24 conditions of $(\alpha,\beta,\eta,r,l,g)$ in Ref.~\cite{13Takeda}
are calculated to give an average value of $0.98 \pm 0.01$.
This value is calculated using only $\ket{j,k}\bra{m,n}$ elements for $j+k=m+n$
on the assumption that the non-zero $\ket{j,k}\bra{m,n}$ elements for $j+k\neq m+n$ in the experimental $\hat{\rho}_\text{out}$
are attributed to the imperfect measurement scheme~\cite{13Takeda}
(otherwise the fidelity drops to $0.92 \pm 0.04$).
The effect of two-photon input states of around 6{\%} are also taken into account (see Appendix C).
This high fidelity demonstrates that our theory accurately models the experiment.

%%%%%%%%%%%%%%%%% Entanglement teleportation %%%%%%%%%%%%%%%%%%%%%%%%%%%%%%%%%%%%%%%%%%%%%%%%%%%%%%%%%%%%%%%%%%%%%

\section{Teleportation of DV entanglement}\label{sec:swapping}

\begin{figure}[!ht]
\begin{center}
\includegraphics[scale=0.6,clip]{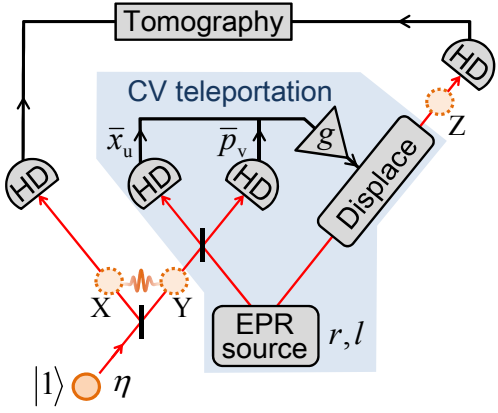}
\end{center}
\caption{(color online) Schematic of hybrid entanglement swapping.}
\label{fig:swapping_scheme_ps}
\end{figure}

In a typical teleportation experiment,
a specific set of quantum states, which is known to the experimentalists, is prepared and used as test states for the teleporter.
However, the genuine quantum nature of teleportation lies in the fact that
it can teleport arbitrary unknown quantum states.
Such quantum nature can manifest itself directly when one half of an entangled pair is teleported.
In this case, for example, a qubit that is to be teleported is indeed in a completely random state on its own while being maximally entangled with another qubit.
The entanglement is then transferred via teleportation, thus revealing the quantum nature of teleportation.
Such teleportation of entanglement, usually referred to as entanglement swapping,
has been proposed and experimentally realized separately
in DV~\cite{93Zukowki,98Pan,02Sciarriono} and CV~\cite{99Tan,99vanLoock,04Jia,05Takei} optical systems.
Here an interesting question arises for our hybrid system:
can CV teleportation transfer DV entanglement?
The original proposal of this hybrid entanglement swapping was made in Ref.~\cite{99Polkinghorne}
as CV teleportation of polarization entanglement between photons.
The condition for violating the Clauser-Horne-type inequality by photon-counting measurement was discussed,
but its experimental demonstration has not been reported yet.

Here we propose another type of hybrid entanglement-swapping experiment
in a readily implementable form.
The main difference from the original proposal in Ref.~\cite{99Polkinghorne} is that
the polarization-entangled photons and photon-counting measurements
are replaced by a single photon split at a beam splitter and homodyne measurements, respectively.
The demonstration of our proposal will give a distinct proof of the fact
that CV teleportation operates non-classically on the DV subspace of $\{\ket{0},\ket{1}\}$.
In addition, this hybrid scheme allows for a more efficient transfer of DV entanglement than previous DV schemes~\cite{93Zukowki,98Pan,02Sciarriono},
and thus has applications in practical quantum communication~\cite{98Briegel,01Duan}.

Our proposal is illustrated in Fig.~\ref{fig:swapping_scheme_ps}.
At the first stage, a heralded single photon is generated based on the method in Refs.~\cite{12Takeda,07Nielsen}.
The photon incident on a 50:50 beam splitter yields maximally entangled single-rail qubits
$\ket{\psi}_\text{XY}=(\ket{0}_\text{X}\ket{1}_\text{Y}+\ket{1}_\text{X}\ket{0}_\text{Y})/\sqrt{2}$.
The qubit of mode Y is then teleported to mode Z via the CV teleporter in Ref.~\cite{13Takeda}.
The two-mode density matrix of modes X and Z can be obtained by
the homodyne-tomography method in Ref.~\cite{04Babichev}.
Finally, entanglement in the final state can be assessed by a violation of the positivity after partial transposition~\cite{96Pere}.
Below we model the proposed experiment in the transition-operator formalism,
and derive a sufficient experimental condition to observe the entanglement after teleportation.
We also deduce the optimal gain to maximize the transferred entanglement, which is quantified by the logarithmic negativity~\cite{02Vidal}.

\begin{figure*}[!t]
\centering
\subfigure[]{
\hspace{1mm}\includegraphics[clip,scale=0.3]{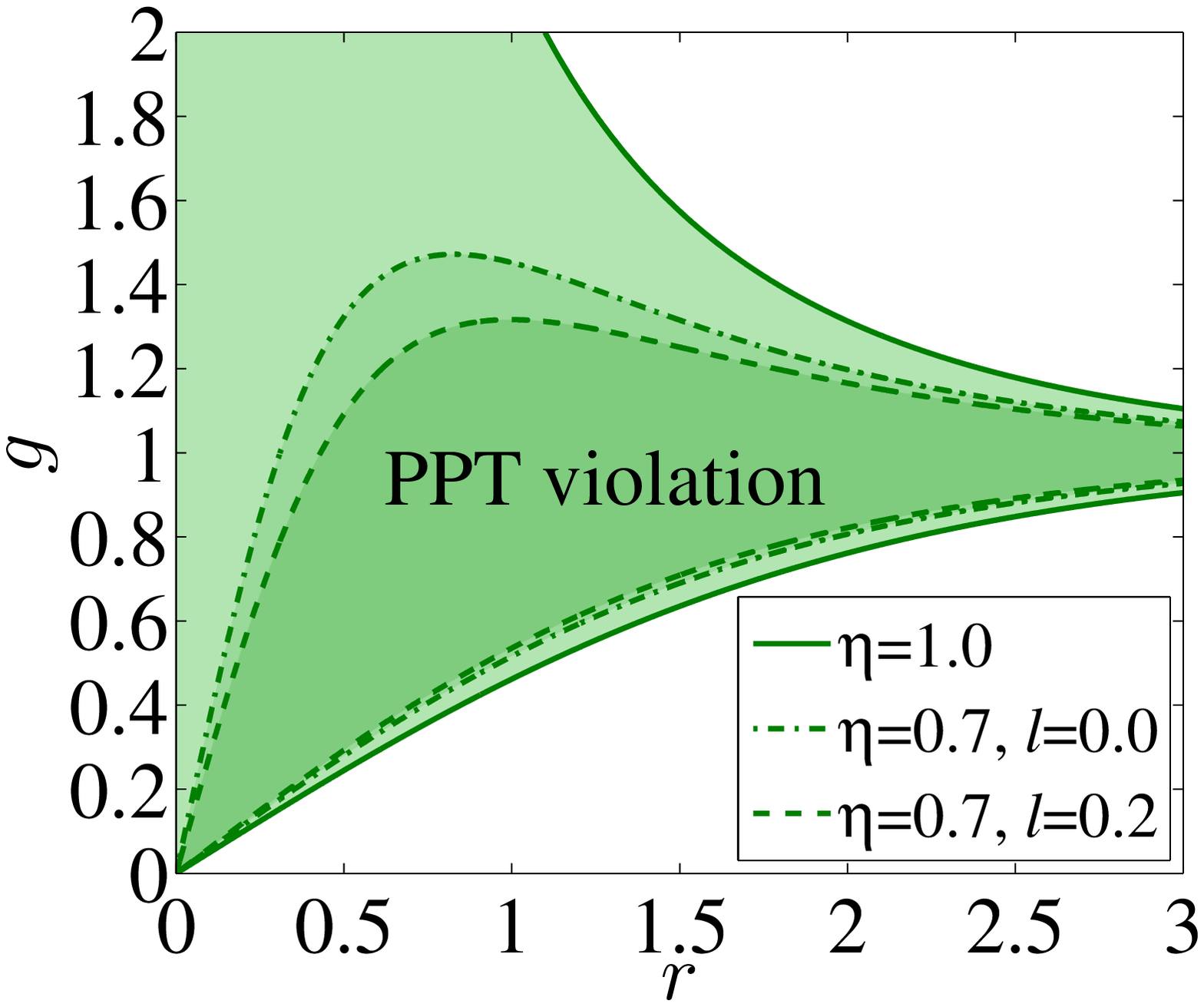}\hspace{1mm}
\label{sfig:entangle_condition_ps}
}
\subfigure[]{
\hspace{1mm}\includegraphics[clip,scale=0.3]{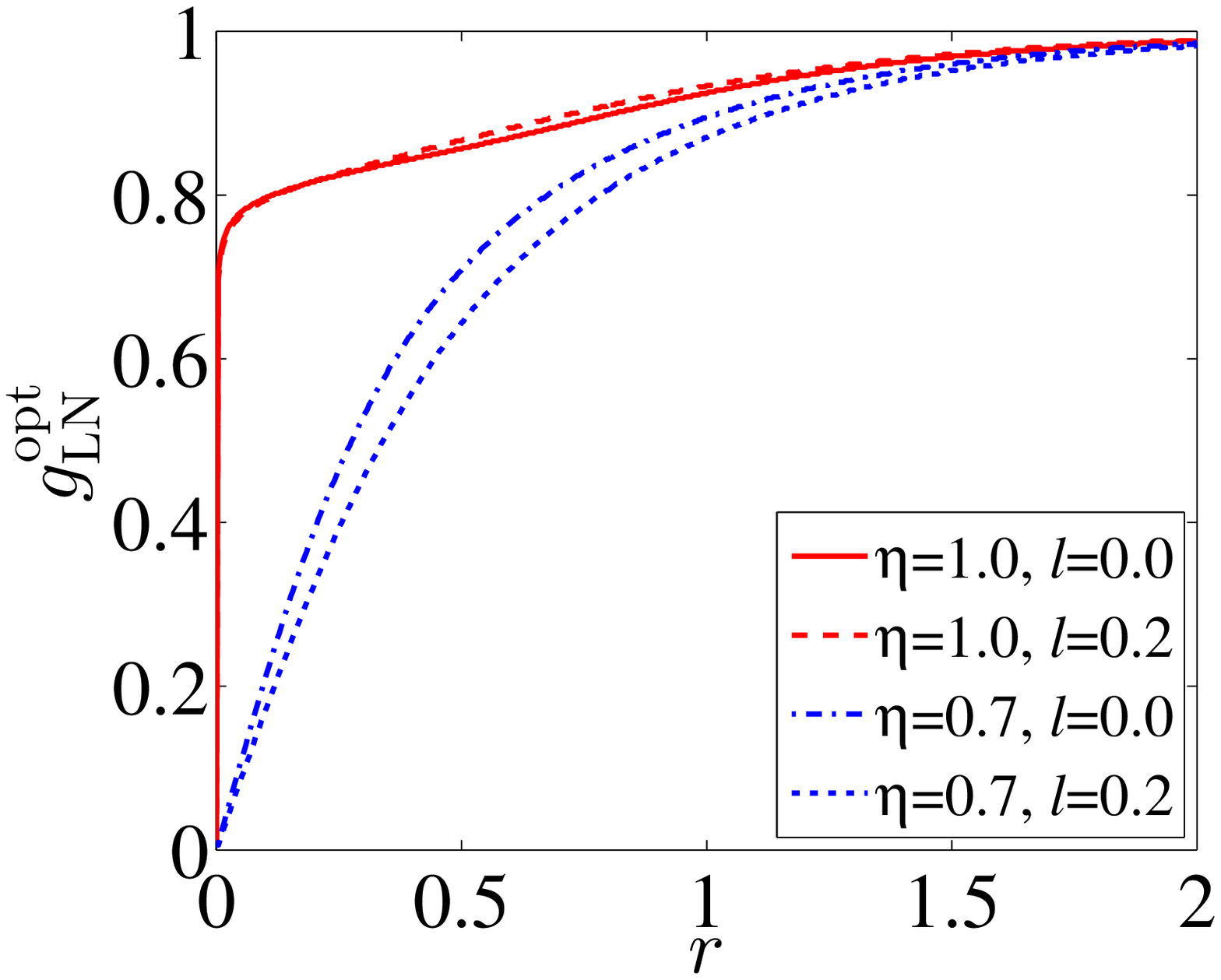}\hspace{1mm}
\label{sfig:log_neg_g_opt_ps}
}
\subfigure[]{
\hspace{1mm}\includegraphics[clip,scale=0.3]{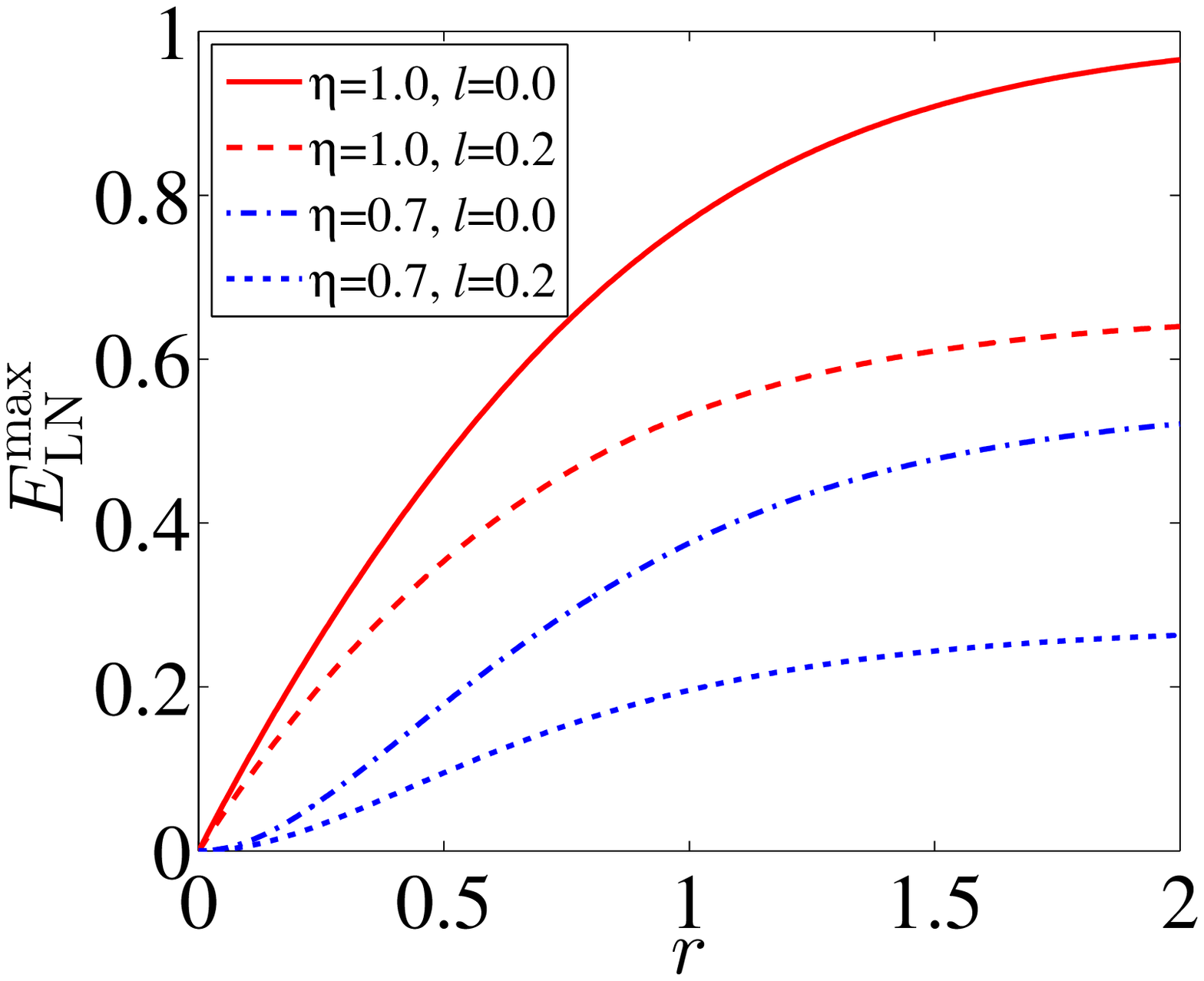}\hspace{1mm}
\label{sfig:log_neg_g_opt_log_neg_ps}
}
\subfigure[]{
\hspace{1mm}\includegraphics[clip,scale=0.30]{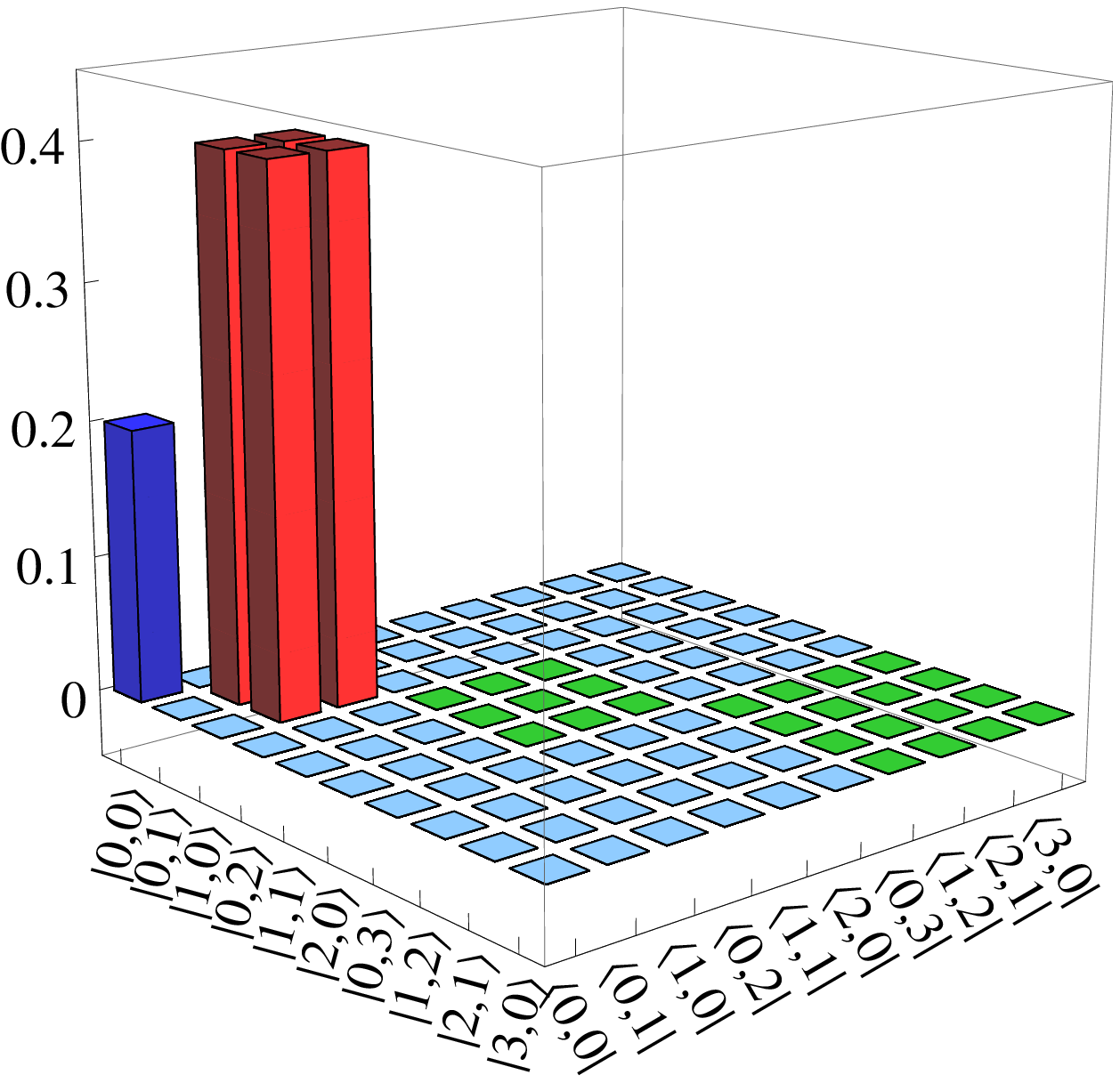}\hspace{1mm}
\label{sfig:swapping_input}
}
\subfigure[]{
\hspace{1mm}\includegraphics[clip,scale=0.30]{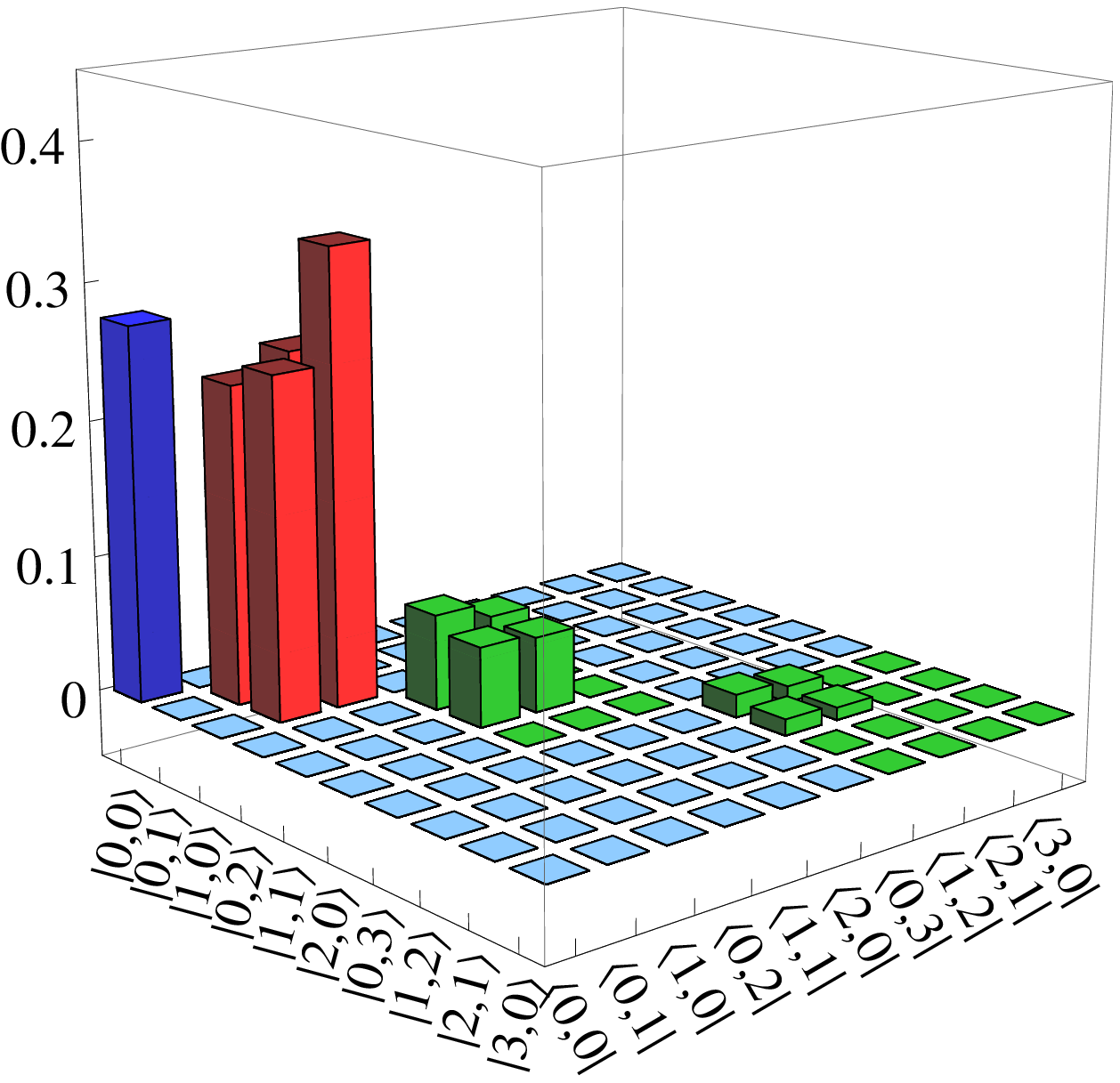}\hspace{1mm}
\label{sfig:swapping_output}
}
\caption{(color online) Simulation results of entanglement swapping.
(a) Area of $(r,g)$ where PPT of $\hat{\rho}_\text{XZ}$ is violated is filled with green [Eq.~(\ref{entanglement_condition})].
(b) $r$ dependence of $g_\text{LN}^\text{opt}$.
(c) $r$ dependence of $E_\text{LN}^\text{opt}(\hat{\rho}_\text{XZ})$.
LN of the initial state is $E_\text{LN}^\text{max}(\hat{\rho}_\text{XY})=1$ for $\eta=1.0$ and $E_\text{LN}^\text{max}(\hat{\rho}_\text{XY})=0.548$ for $0.7$.
(d) Theoretical $\hat \rho_\text{XY}$ in Eq.~(\ref{input_density_swap}) for $\eta=0.8$ (real part).
(e) Theoretical $\hat \rho_\text{XZ}$ in Eq.~(\ref{output_density_swap}) for $(\eta,r,l,g)=(0.8,1.0,0.2,0.90)$ (real part).
Note that there is no imaginary component for both $\hat \rho_\text{XY}$ and  $\hat \rho_\text{XZ}$.
}
\label{fig:entangle_swapping_cond_log_neg}
\end{figure*}

Experimentally, the initial single-photon state becomes a state mixed with vacuum corresponding to a loss fraction of $1-\eta$.
The entangled state $\hat \rho _\text{XY}$ after the 50:50 beam splitter can thus be modeled by
\begin{align}
&\hat \rho_\text{XY}\!=\!(1-\eta )\ket{0}_\text{X}\!\bra{0} \!\otimes\! \ket{0}_\text{Y}\!\bra{0}
+\eta\ket{\psi}_\text{XY}\!\bra{\psi}.
\label{input_density_swap}
\end{align}
CV teleportation from mode Y to Z replaces
$\ket{m}_\text{Y}\!\bra{n}$ in Eq.~(\ref{input_density_swap})
by its corresponding transition operator $\hat{T}_\text{Z}^{mn}\equiv\hat{T}_\text{Z}(\ket{m}_\text{Y}\!\bra{n})$ ($m,n=0,1$).
The final density matrix is thus written as
\begin{align}
\hat \rho_\text{XZ}&=(1-\eta )\ket{0}_\text{X}\!\bra{0} \otimes \hat{T}_\text{Z}^{00}
+\!\frac{\eta}{2}\Bigl(\ket{1}_\text{X}\!\bra{1}\!\otimes\!\hat{T}_\text{Z}^{00} \nonumber\\
&+\!\ket{1}_\text{X}\!\bra{0}\!\otimes\!\hat{T}_\text{Z}^{01}
\!+\!\ket{0}_\text{X}\!\bra{1}\!\otimes\!\hat{T}_\text{Z}^{10}
\!+\!\ket{0}_\text{X}\!\bra{0}\!\otimes\!\hat{T}_\text{Z}^{11}\Bigr).
\label{output_density_swap}
\end{align}
The success of entanglement swapping can be confirmed by witnessing entanglement in $\hat \rho_\text{XZ}$.
It is known that the positivity of the partial transposition (PPT) of $\hat \rho_\text{XZ}$
(namely, partial transposition for just one subsystem, X or Z)
is a necessary condition for $\hat \rho_\text{XZ}$ to be separable~\cite{96Pere}.
The partial transposition of Eq.~(\ref{output_density_swap}) for mode X is
\begin{align}
\hat \rho_\text{XZ}^{\text{T}_\text{X}}
&=(1-\eta )\ket{0}_\text{X}\!\bra{0} \otimes \hat{T}_\text{Z}^{00}
+\!\frac{\eta}{2}(\ket{1}_\text{X}\!\bra{1}\!\otimes\!\hat{T}_\text{Z}^{00} \nonumber\\
&+\!\ket{0}_\text{X}\!\bra{1}\!\otimes\!\hat{T}_\text{Z}^{01}
\!+\!\ket{1}_\text{X}\!\bra{0}\!\otimes\!\hat{T}_\text{Z}^{10}
\!+\!\ket{0}_\text{X}\!\bra{0}\!\otimes\!\hat{T}_\text{Z}^{11})\nonumber\\
&=  \sum_{k=-1}^\infty \hat \rho_{k},
\end{align}
where $\hat \rho_k$ ($k\ge-1$) is defined as
\begin{align}
\hat \rho_k =&  a_{k} \ket{0}_\text{X} \bra{0} \otimes \ket{k}_\text{Z} \bra{k}
+b_{k} \ket{0}_\text{X} \bra{1} \otimes \ket{k}_\text{Z} \bra{k+1} \nonumber  \\
&\quad+b_{k} \ket{1}_\text{X} \bra{0} \otimes \ket{k+1}_\text{Z} \bra{k} \nonumber \\
&\quad\quad+c_{k} \ket{1}_\text{X} \bra{1} \otimes \ket{k+1}_\text{Z} \bra{k+1}
\end{align}
with coefficients $a_k,b_k,c_k\ge0$, given by
\begin{align}
a_{k} &= (1-\eta ) T_{00\to kk} + \frac{\eta}{2}T_{11\to kk}\> (k \ge 0), \>\> 0 \> (k=-1),  \nonumber \\
b_{k} &= \frac{\eta}{2} T_{10\to k\!+\!1\> k}\> (k \ge 0), \>\> 0 \> (k=-1),    \nonumber \\
c_{k} &= \frac{\eta}{2} T_{00\to k\!+\!1\>k\!+\!1}\ (k \ge -1).
\label{abcd}
\end{align}
Here $\hat \rho_m \hat \rho_n = \hat O$ can be proven for all $m$ and $n$ ($m,n\ge -1$, $m \neq n$).
This means that the set of eigenvalues of $\hat \rho^{\text{T}_\text{X}}_\text{XZ}$ is equivalent to
the sum of sets of eigenvalues of each $\hat \rho_k$ ($k\ge -1$).
When at least one of $\hat \rho_k$ has a negative eigenvalue,
$\hat \rho^{\text{T}_\text{X}}_\text{XZ}$ fails to be positive,
in which case we know that $\hat \rho_\text{XZ}$ is entangled.
Since $\hat \rho_{-1}$ is always positive,
the sign of the following eigenvalues $\lambda_k^{\pm}$ of each $\hat \rho_k$ ($k \ge 0$) determines the positivity,
\begin{equation}
\lambda_k^{\pm} = \frac{a_k + c_k \pm \sqrt{(a_k - c_k)^2 + 4 b_k^2}}{2}.
\label{eq:eigenvalue_rhok}
\end{equation}
Since $\lambda_k^+>0$,
we focus on the condition $\lambda^-_k<0$.
From Eqs.~(\ref{photon_dis_00}), (\ref{photon_dis_10}), (\ref{photon_dis_11}) and (\ref{abcd}),
this condition can be calculated as
\begin{align}
\frac{3\eta-2}{2-\eta}> 2\tau -\frac{1}{g^2},
\label{entanglement_condition}
\end{align}
with $\tau$ defined  by Eq.~(\ref{gamma}).
This is the necessary and sufficient condition for $\hat \rho_k$ to have a negative eigenvalue.
Since  Eq.~(\ref{entanglement_condition}) is independent of $k$,
this inequality is also the necessary and sufficient condition
for $\hat \rho^{\text{T}_\text{X}}_\text{XZ}$ to have at least one negative eigenvalue.
In other words, when the experimental setting $(\eta,r,l,g)$ satisfies Eq.~(\ref{entanglement_condition}),
entanglement in $\hat \rho_\text{XZ}$ can be verified by a violation of the PPT criterion.
When $g$ can be chosen arbitrarily,
Eq.~(\ref{entanglement_condition}) is most easily satisfied
by choosing $g^\text{opt}_\text{PPT}=\tanh r$, when the right hand side takes its minimum value $2l-1$.
At this gain, Eq.~(\ref{entanglement_condition}) is reduced to
\begin{equation}
\frac{\eta}{2-\eta}  > l,  \label{entanglement_condition_eta_l}
\end{equation}
independent of the parameter $r$.
Therefore $\eta$ and $l$ are the critical parameters in this experiment.
As long as $\eta$ and $l$ satisfy Eq.~(\ref{entanglement_condition_eta_l}),
PPT of $\hat{\rho}_\text{XZ}$ can be violated for any non-zero $r$ at the optimal gain $g^\text{opt}_\text{PPT}=\tanh r(=g_\text{att})$.
The range of $(r,g)$ satisfying Eq.~(\ref{entanglement_condition}) is plotted in Fig.~\ref{sfig:entangle_condition_ps} for various $(\eta,l)$ values.
For $\eta=1$, Eq.~(\ref{entanglement_condition}) gives
\begin{align}
\frac{\cosh r-1}{\sinh r}<g<\frac{\cosh r+1}{\sinh r},
\end{align}
and the range does not depend on $l$.
For $\eta<1$, the violation range gets smaller with an increasing $l$,
and finally vanishes when Eq.~(\ref{entanglement_condition_eta_l}) is violated.

In addition, the degree of entanglement in $\hat \rho_\text{XZ}$ can be assessed by the logarithmic negativity (LN),
given by~\cite{02Vidal}
\begin{equation}
E_\text{LN} ( \hat \rho_\text{XZ} )\!\equiv\!\log_2\! || \hat \rho_\text{XZ}^{\text{T}_\text{X}}  ||
\!=\!\log_2\!\left[1\!+\!\sum_k(|\lambda_k^-|\!-\!\lambda_k^-)\right],
\label{eq:def_lognega}
\end{equation}
where $||\hat \rho || = \text{Tr}(\hat \rho ^\dagger \hat \rho)^{1/2}$.
This quantifies the degree to which $\hat \rho_\text{XZ}^{\text{T}_\text{X}}$ fails to be positive;
the state is entangled when $E_\text{LN}( \hat \rho_\text{XZ} )>0$,
and a maximally entangled state gives $E_\text{LN}( \ket{\psi}_\text{XY}\!\bra{\psi})=1$.
The value of $E_\text{LN} ( \hat \rho_\text{XZ} )$ can be directly calculated
by inserting $\lambda_k^-$ of Eq.~(\ref{eq:eigenvalue_rhok}) into Eq.~(\ref{eq:def_lognega}).
Figures~\ref{sfig:log_neg_g_opt_ps} and \ref{sfig:log_neg_g_opt_log_neg_ps} each show the $r$ dependence of
the optimal gain $g_\text{LN}^\text{opt}$ that maximizes $E_\text{LN}$ and the maximum value $E_\text{LN}^\text{max}$ for various $(\eta,l)$ values.
$g_\text{LN}^\text{opt}$ is always larger than $g^\text{opt}_\text{PPT}=\tanh r$.
By increasing $r$, $E_\text{LN}^\text{max}$ monotonically increases and approaches the LN of the initial state,
but the upper bound is limited by loss $l$.

Finally, the viability of this experiment is examined based on current technologies.
When $l=0.2$ is assumed as reported in Ref.~\cite{13Takeda},
Eq.~(\ref{entanglement_condition_eta_l}) requires an input efficiency of $\eta>1/3$.
Since heralded single-photon sources with efficiency up to $\eta\sim0.8$ have been already reported~\cite{13Miwa},
the verification of entanglement is possible.
As an example, for $(\eta,r,l)=(0.8,1.0,0.2)$,
the maximum LN of $E_\text{LN}^\text{max}(\hat\rho_\text{XZ})=0.30$ is obtained at $g_\text{LN}^\text{opt}=0.90$ after teleportation
for an initial state with LN $E_\text{LN}(\hat\rho_\text{XZ})=0.70$.
The theoretical density matrices $\hat \rho_\text{XY}$ and $\hat \rho_\text{XZ}$ under this condition
are illustrated in Fig.~\ref{sfig:swapping_input} and \ref{sfig:swapping_output}.
It can be seen that the original DV entanglement,
which manifests itself as four elements in the subspace $\{\ket{0}_\text{X}\ket{1}_\text{Y}, \ket{1}_\text{X}\ket{0}_\text{Y}\}$,
is transformed into the other elements in the subspace $\{\ket{0}_\text{X}\ket{k+1}_\text{Z}, \ket{1}_\text{X}\ket{k}_\text{Z}\}$ ($k\ge0$)
due to the thermalization effect on mode Y in teleportation.

%%%%%%%%%%%%%%%%% Conclusion %%%%%%%%%%%%%%%%%%%%%%%%%%%%%%%%%%%%%%%%%%%%%%%%%%%%%%%%%%%%%%%%%%%%%

\section{Conclusion}\label{sec:conclusion}

We have developed a general formalism to describe the transformation of DV states by a CV teleportation channel.
This formalism can be used to model various hybrid teleportation experiments and investigate the optimal gain tuning for a given figure of merit.
The key element of our formalism is a transition operator $\hat{T}$,
which describes how each density matrix element is transformed by the CV teleportation channel.
This operator includes two parameters $(\tau,g)$ that characterize the channel properties.
By appropriately choosing $(\tau,g)$, we can describe experimental imperfections such as loss on the input state and impurity of squeezing,
and we can also analyze more complex teleportation channels composed of consecutive CV teleportation, pure attenuation and amplification.

We have applied our formalism to CV teleportation of a dual-rail qubit,
and discussed the optimal gain to obtain the maximum fidelity under various experimental conditions.
The validity of our model is confirmed by the good agreement between our theoretical prediction and the actual experimental results in Ref.~\cite{13Takeda}.
We have also proposed and modeled CV teleportation of DV entanglement in the form of a split single photon.
It has been proven that, provided the efficiency of the input qubits and the loss on the resource squeezing satisfy a certain condition, DV entanglement can be teleported for any non-zero squeezing level by optimally tuning the gain.
This experiment can be thus readily realized with current technology.

%%%%%%%%%%%%%%%%% Appendix %%%%%%%%%%%%%%%%%%%%%%%%%%%%%%%%%%%%%%%%%%%%%%%%%%%%%%%%%%%%%%%%%%%%%

\section*{ACKNOWLEDGEMENTS}

This work was partly supported by the SCOPE program of the MIC of Japan, PDIS, GIA, G-COE, and APSA commissioned by the MEXT of Japan,
FIRST initiated by the CSTP of Japan, and ASCR-JSPS.
S.T. and M.F. acknowledge financial support from ALPS.
P.v.L. acknowledges support from the BMBF in Germany through QuOReP and HIPERCOM.

\appendix

\section*{APPENDIX A: COMMUTATION OF BEAM-SPLITTER TRANSFORMATION AND TELEPORTATION PROCESS}\label{sec:commu_U_2tele}

Here we prove Eq.~(\ref{eq:commutation_TU})
by deriving the Wigner functions corresponding to the density matrices of both left and right hand sides of Eq.~(\ref{eq:commutation_TU}).
For this purpose we define the Wigner function of $\hat \rho_\text{XY}$ as $W_\text{XY}(\xi_\text{X},\xi_\text{Y})$.
On the right hand side of Eq.~(\ref{eq:commutation_TU}),
$\hat{T}_\text{XY}(\cdot)$ is first applied to the input state.
By the definition of $\hat{T}_\text{XY}$ and Eq.~(\ref{imperfect_EPR_GC}),
this process transforms the Wigner function into
\begin{align}
\frac{1}{g^4}\!\iint d\xi''_\text{X}d\xi''_\text{Y} W_\text{XY}(\xi''_\text{X} ,\xi''_\text{Y} )
G_{\tau}\!\left(\!\frac{\xi_\text{X}}{g}\!-\!\xi''_\text{X}\!\right)
G_{\tau}\!\left(\!\frac{\xi_\text{Y}}{g}\!-\!\xi''_\text{Y}\!\right).
\label{eq:2modeW_tele}
\end{align}
The beam-splitter transformation $\hat{U}$ is then applied
($\hat U\hat a_\text{X}^\dagger \hat U^\dagger=\beta^* \hat a_\text{X}^\dagger - \alpha \hat a_\text{Y}^\dagger$,
$\hat U\hat a_\text{Y}^\dagger \hat U^\dagger = \alpha^* \hat a_\text{X}^\dagger + \beta \hat a_\text{Y}^\dagger$).
This process transforms the arguments $\xi_\text{X}$ and $\xi_\text{Y}$ in Eq.~(\ref{eq:2modeW_tele}) as follows,
\begin{align}
&\frac{1}{g^4}\!\iint d\xi''_\text{X}d\xi''_\text{Y} W_\text{XY}(\xi''_\text{X} ,\xi''_\text{Y} )\nonumber\\
&\times G_{\tau}\!\left(\!\frac{\beta^* \xi_\text{X}-\alpha \xi_\text{Y}}{g}\!-\!\xi''_\text{X}\!\right)
G_{\tau}\!\left(\!\frac{\alpha^* \xi_\text{X}+\beta \xi_\text{Y}}{g}\!-\!\xi''_\text{Y}\!\right).
\label{eq:commutation_RHS}
\end{align}
On the left hand side of Eq.~(\ref{eq:commutation_TU}),
the input state is first transformed into
$W_\text{XY}(\beta^* \xi_\text{X}-\alpha\xi_\text{Y}, \alpha^* \xi_\text{X}+\beta\xi_\text{Y})$
by the operator $\hat{U}$, which is then transformed into
\begin{align}
&\frac{1}{g^4}\!\iint d\xi'_\text{X} d\xi'_\text{Y} W_\text{XY} (\beta^* \xi'_\text{X}\!-\!\alpha \xi'_\text{Y},\alpha^* \xi'_\text{X}\!+\!\beta \xi'_\text{Y} ) \nonumber \\
&\quad\times\!G_{\tau}\!\left(\!\frac{\xi_\text{X}}{g}\!-\!\xi'_\text{X} \!\right)G_{\tau}\!\left(\!\frac{\xi_\text{Y}}{g}\!-\!\xi'_\text{Y} \!\right)
\label{eq:commutation_LHS}
\end{align}
by the teleportation process.
Equation (\ref{eq:commutation_LHS}) becomes equal to Eq.~(\ref{eq:commutation_RHS})
by replacing the integration variable $(\xi'_\text{X},\xi'_\text{Y})$ in Eq.~(\ref{eq:commutation_LHS})
by $(\xi''_\text{X},\xi''_\text{Y})=(\beta^* \xi'_\text{X} - \alpha \xi'_\text{Y},\alpha^* \xi'_\text{X} + \beta \xi'_\text{Y})$.
Thus, the Wigner functions of the left and right hand sides in Eq.~(\ref{eq:commutation_TU}) are equivalent.
This fact proves Eq.~(\ref{eq:commutation_TU}).

\section*{APPENDIX B: EFFECT OF MEASUREMENT INEFFICIENCIES}\label{sec:measurement_ineff}

Here we show how to take into account
the measurement inefficiencies at the detectors for modes u, v, and output in Fig.~\ref{fig:schematic_teleportation}
when they are no longer negligible.
In conclusion, these measurement inefficiencies can be incorporated into the input efficiency ($\eta$) and the loss on the EPR state ($l$)
on the assumption that all these detector efficiencies are the same ($\eta_\text{d}$).
We use the following lemma to prove this fact:
symmetric losses on two modes after a beam splitter is equivalent to the same amount of
symmetric losses on two modes before their combination.
First, the measurement loss $1-\eta_\text{d}$ in modes u and v can be incorporated in the input loss $1-\eta$ and the loss on mode A.
Next, the measurement loss $1-\eta_\text{d}$ on the output state adds loss to mode B as well as to the amplitude of displacement.
The losses on modes A, B can be incorporated in the initial loss $l$.
The loss on the displacement amplitude attenuates the effective feedforward gain from $g$ to $\sqrt{\eta_\text{d}}g$.
In our analysis in Secs.~\ref{sec:qubit_tele} and \ref{sec:swapping},
all these measurement inefficiencies are included in $\eta$ and $l$.
The gain reduction is neglected,
but this effect only changes the scales of the gain axes in Figs.~\ref{fig:tele_simulation_results} and \ref{fig:entangle_swapping_cond_log_neg}.

\section*{APPENDIX C: EFFECT OF TWO-PHOTON INPUT STATE}\label{sec:multiphoton_tele}

In the simulation of Sec.~\ref{sec:qubit_tele},
we have assumed that the initial dual-rail qubit has no multi-photon terms.
However,  the experimental qubit state usually has non-zero multi-photon terms as in Ref.~\cite{13Takeda}.
The input state in  Ref.~\cite{13Takeda} has around 6\% of two-photon terms in the subspace of $\{\ket{0,2},\ket{1,1},\ket{2,0}\}$,
which cannot be neglected.
This effect can be taken into account by the following method.

Extending Eq.~(\ref{eq:decomposed_input}),
we can assume that the input state with two-photon terms is modeled by
\begin{align}
\hat \rho_\text{in}&= \hat U^\dagger\left(\hat \rho_\text{X} \otimes \ket{0}_\text{Y}\!\bra{0}\right)\hat U, \nonumber \\
\hat \rho_\text{X} &=   (1\!-\!\eta_1\!-\!\eta_2 ) \ket{0}_\text{X}\!\bra{0} + \eta_1 \ket{1}_\text{X}\!\bra{1} + \eta_2 \ket{2}_\text{X}\!\bra{2}. \label{input_unitary_mod}
\end{align}
By appropriately choosing $\eta_1$ and $\eta_2$, this $\hat\rho_\text{in}$ simulates well the experimental input state,
yielding an average fidelity of $0.98\pm0.01$ for the six input states in Ref.~\cite{13Takeda}.
In the transition-operator formalism, the teleported state for this $\hat{\rho}_\text{in}$ is given by
\begin{align}
\hat \rho_\text{out}
\!=\!\hat{U}^\dagger \left\{\left[\eta_2\hat{T}_\text{X}^{22}\!+\!\eta_1\hat{T}_\text{X}^{11}\!+\!(1\!-\!\eta_1\!-\!\eta_2)\hat{T}_\text{X}^{00}\right]
\!\otimes\!\hat{T}_\text{Y}^{00}\right\}\hat U,
\end{align}
where  $\hat{T}_\text{X}^{22}\equiv\hat{T}_\text{X}(\ket{2}_\text{X}\!\bra{2})$.
In order to calculate each element of $\hat{\rho}_\text{out}$,
the expression of the coefficient $T_{22\to jk}$ of $\hat{T}_\text{X}^{22}$ is needed
in addition to Eqs.~(\ref{photon_dis_00}) and (\ref{photon_dis_11}).
This can be calculated from Eqs.~(\ref{imperfect_EPR_GC}), (\ref{gamma}), and (\ref{photon_num_wigner}) as
\begin{align}
T_{22\to jk}&=\frac{2(\lambda\!-\!1)^{k-2}}{(\lambda\!+\!1)^{k+3}}
\Big[\left(\lambda\!-\!2g^2\!+\!1 \right)^2\!(\lambda\!-\!1)^2 \nonumber\\
&+\!8kg^2(\lambda^2\!-\!2g^2\lambda\!+\!g^2\!-\!1)\!+\!8k^2g^4 \Big] \delta_{j,k}. \label{photon_dis_22}
\end{align}
Using the results above, the two-photon effect of the input state
can be considered for the derivation of
the density matrices ($\hat{\rho}_\text{in}$, $\hat{\rho}_\text{out}$)
and the fidelity ($F_\text{state}$, $F_\text{qubit}$).

%%%%%%%%%%%%%%%%%%%%%%%%%%%%%%%%%%%%%%%%%%%%%%%%%%%%%%%%%%%%%%%%%%%%%%%%%%%%%%%%%

\end{document}